\documentclass[journal=jpccck,manuscript=article]{achemso}
\usepackage{mathrsfs}
\usepackage{amsmath}
\usepackage{amsmath}
\usepackage{graphicx}
\usepackage{dcolumn}
\usepackage{bm}
\newcommand{\bpsi}{\boldsymbol{\psi}}
\newcommand{\bsigma}{\boldsymbol{\sigma}}

\title{Resonance electronic excitation energy transfer in the quantum dot system}

\author{O. P. Chikalova-Luzina}
\email{o_chikalova@mail.ru}
\author{D. M. Samosvat}
\email{samosvat@yandex.ru}
\affiliation{Ioffe Institute, 26 Politekhnicheskaya, St Petersburg 194021, Russian Federation}
\author{V. M. Vyatkin}
\affiliation{St. Petersburg State Electrotechnical University, St. Petersburg, 197376, Russia}
\author{G. G. Zegrya}
\affiliation{Ioffe Institute, 26 Politekhnicheskaya, St Petersburg 194021, Russian Federation}

\begin{document}
\begin{abstract}
Microscopic theory of the nonradiative energy transfer in a system of III-V semiconductor quantum dots is elaborated in our work. The energy transfer both due to direct Coulomb and due to exchange interactions between two quantum dots (energy donor and acceptor) is considered.  An analysis of energy transfer process is performed in the frame of the Kane model that provides the most adequate description of the real energy spectra and wave functions of III-V semiconductors.
The density-matrix method is applied, which enabled us to analyze the energy transfer rate both in the weak-interaction approximation and in the strong-interaction approximation. For the first time the detailed analytical calculations of the exchange energy transfer rate for the quantum dot system are performed. The analytical expressions for contributions to the transfer rate are derived.  The numerical calculations showed that at nearly contact distances between two quantum dots the rate of the energy transfer due to the direct Coulomb interaction as well as by exchange interaction can reach the saturation.  At the small distances, these two contributions can be of the same order and can have the same value in the saturation range. It is revealed that the exchange interaction should be taken into consideration in qualitative describing the energy transfer at small distances between the quantum dot donor and the quantum dot acceptor.
\end{abstract}
\maketitle


\section{Introduction}
The electronic excitation energy transfer between quantum systems is one of the most important fundamental problems of modern physics \cite{ref01}.
The phenomenon consists in that the electronic excitation energy is transferred from an energy donor [atom, molecule, semiconductor quantum
dot (QD) or quantum well (QW)] to an energy acceptor. The following energy transfer mechanisms are distinguished: well-known radiative mechanism
(when the donor emits a photon and the acceptor it then absorbs (see, e.g., \cite{ref02,ref03}), nonradiative mechanism (in which energy is transferred from
a donor to an acceptor via a single-step mechanism in contrast to the radiative energy transfer) \cite{ref04,ref05}, and electron transfer mechanism (when
an excited electron is transferred from the energy donor to the acceptor) \cite{ref06}. The last two mechanisms quench the luminescence from a donor,
but the first of these leads to a sensitized fluorescence of the acceptor, and the second yields a positively charged donor and a negatively
charged acceptor (pairs of ions in the case of molecules). These mechanisms are fundamentally different: the nonradiative energy transfer occurs
due to the Coulomb interaction of electrons of the energy donor and acceptor, whereas the electron transfer is only determined by the overlapping
of the wave functions of the corresponding states of the donor and acceptor.

The nonradiative energy transfer was first observed in 1923 in
experiments on the sensitized fluorescence of atoms in the gas phase \cite{ref07}. Later, experiments of this kind were performed for molecule vapors
\cite{ref08}, liquid solutions of dyes \cite{ref09,ref10,ref11}, and solid solutions of organic molecules \cite{ref12}. In parallel, a multitude of studies revealed the part
played by the nonradiative energy transfer in biological systems (photosynthesis, in particular) \cite{ref13} (see also references in \cite{ref14}).
Subsequently, the method based on the energy transfer between molecules of organic dyes found wide use in biological and medical experiments
(see, e.g., \cite{ref15,ref16}).

In systems including semiconductor QDs, the nonradiative energy transfer was first observed in 1996 \cite{ref17} and then started to be intensively
studied both experimentally \cite{ref18} and theoretically \cite{ref19,ref20,ref21,ref22}. The interest is primarily due to the fact that the application of QDs has extended
the potential of biological and medical experiments, both {\it in vivo} and {\it in vitro}, owing to their unique optical properties (narrow luminescence spectra,
possibility of changing the spectral characteristics of QDs by varying their size due to the quantum confinement effect). Together with the optical
characteristics, QDs have advantages for the photostability and chemical stability over organic dyes conventionally used in this area of research.
The QD-based resonance energy transfer and its growing application in biology are overviewed in
\cite{ref23}. Representative examples of biological
experiments based on the energy transfer between QDs are presented in review \cite{ref24}.
With particular interest in biosensing application, the foundational and theoretical works on the energy transfer between QDs are discussed in
overview \cite{ref25}. The extensive review on energy transfer with semiconductor quantum
dot bioconjugares is presented in \cite{ref26}.
Resent papers exhibited in these reviewers demonstrate that an investigation of nonradiative resonance energy transfer remains the topical problem until the preset time.

The possibility of technical applications of the mechanism of nonradiative energy transfer between QDs for
constructing fast quantum computers \cite{ref27}, QD semiconductor lasers \cite{ref28,ref29}, and solar cells
\cite{ref30} is discussed in the literature, which also stimulates studies of this physical process.
In the development of the optoelectronic devices, the considerable attention is presently given to the
exploiting of the densely packed thin QD films. \cite{add01}
, which needs the consideration of the peculiarities
of the energy transfer at small QDs separation.

The first quantum-mechanical description of the nonradiative energy transfer was developed by Forster for molecular systems \cite{ref04}. He assumed that the
energy transfer mostly occurs as a result of the dipole-dipole interactions between molecules. Then, the theory was extended by Dexter via inclusion
of the dipole-quadrupole and exchange interactions \cite{ref05}. Because the exchange interaction is only determined by the Coulomb interaction and by the
spatial overlapping of wave functions of donor and acceptor carriers, the exchange mechanism allows  energy transfer in those cases when the direct
transfer is forbidden by selection rules. It was shown in \cite{ref05} that the exchange contribution to the nonradiative energy transfer rate may be important
in two cases: (i) when the distance between a donor and an acceptor is short and the wave functions are not strongly localized and (ii) when dipole-dipole
transitions in the acceptor are forbidden by selection rules. In the last case, the exchange contribution becomes of key importance. By the example
of a typical pair of impurities in neighboring lattice cells of a NaCl crystal, a value of $10^{12}$--$10^{13}$ $s^{-1}$ was obtained for the Coulomb contribution
to the energy transfer rate, and a value smaller by one to two orders of magnitude, $10^{10}$--$10^{11}$ $s^{-1}$, for the exchange contribution.

In the Forster--Dexter theory, the nonradiative energy transfer is considered for the case of a "very weak interaction" between an energy donor and
acceptor in terms of the quantum-mechanical perturbation theory. Later, the method of density matrices was suggested, which can analyze various
approximations in the energy-transfer theory and is applicable in the case of a strong interaction between a donor and an acceptor \cite{ref01,ref31}. The
already performed theoretical considerations and experimental studies suggest that the energy transfer between molecules has been sufficiently examined.

The theory of the nonradiative resonance energy transfer in systems including semiconductor quantum structures has been insufficiently developed so far
and is the subject of present-day studies. In \cite{ref32}, the nonradiative resonance energy transfer was first considered for a hybrid nanostructure constituted
by a semiconductor QW and a layer of an organic acceptor. An analysis in the effective mass approximation for describing a Wannier--Mott exciton in a
semiconductor QW and a making macroscopic electrodynamic description of the organic medium demonstrated the high efficiency of the nonradiative
energy transfer from an exciton to an organic molecule, with the possible subsequent emission of light. The authors predicted the possibility
of using hybrid structures of this kind for optical pumping of organic emission sources. Then, the same theoretical approach was used to analyze
the mechanism of the nonradiative resonance energy transfer from a semiconductor QD to an organic matrix \cite{ref33}. It was shown that the transfer of a
considerable part of energy from a QD to the surrounding optically active organic molecules is possible in this mechanism. The authors of this study
noted that this effect will be manifested more clearly under electrical pumping of a QD, compared with the optical pumping. In \cite{ref34,ref35,ref36,ref37},
the theory
of energy transfer in semiconductor--organic medium hybrid structures was further developed. The nonradiative energy transfer in another hybrid
system, QD + protein molecule, was considered in \cite{ref38} by using the density-matrix formalism.

The mechanism of the nonradiative energy transfer between QDs has been studied with the use of various theoretical approaches: tight-binding method \cite{ref19},
method of a semi-empirical pseudopotential \cite{ref20}, and simple effective mass model \cite{ref21,ref22}. It was shown in \cite{ref19,ref20} that the dipole-dipole approximation
of the Coulomb interaction of electrons in a donor QD and an acceptor QD provides an adequate description of the nonradiative energy transfer process in
the case of direct-gap semiconductors, and the dependence of the transfer rate W on the distance d between the QDs is described by a simple law $W \sim 1/d^6$.
The contributions from higher multipoles are negligible down to the contact distances between the donor and acceptor. For indirect-gap semiconductors,
the multipole terms are more important, but the dipole-dipole contributions remain dominant. The authors of \cite{ref21} found that the dipole-dipole contribution
to the energy transfer rate is, as a rule, larger than the dipole-quadrupole contribution. However, the dipole-quadrupole contribution, which depends on
the distance between the QDs as $1/d^8$, should be taken into account in a quantitative description at small distances comparable with the QD size. It
was shown in \cite{ref22} that the dipole-dipole approximation is valid for describing the energy transfer for dipole-allowed transitions in the donor and acceptor
for all distances between the QDs, down to nearly contact distances. It was also demonstrated that the rate of energy transfer from a donor to an acceptor,
which corresponds to dipole-forbidden transitions in the acceptor, is also important and its contribution at nearly contact distances may reach a value
of 25\% relative to the contribution of the transfer for a dipole-allowed transition. When studying the energy transfer between QDs, the authors of \cite{ref22}
neglected the exchange interaction by considering it insignificant and disregarded the admixture of valence-band states to the conduction band states
(neglected the nonparabolicity effect). However, we demonstrated in the study reported here that the contribution of the exchange interaction to the
nonradiative energy transfer should be taken into consideration in the case of nearly contact distances between a donor and an acceptor. It is
noteworthy that the results of \cite{ref19,ref20,ref21,ref22} are in agreement at large donor--acceptor distances and show a significant discrepancy at small distances.
In these studies, the rate of the resonance energy transfer is described within the framework of the first-order perturbation theory.
In \cite{ref39}, the generalized modeling Hamiltonian was used to formulate a model that made it possible to examine the influence exerted by the Stark
effect on the resonance energy transfer between QDs under the action of a short laser pulse. The dynamics variation of the occupation of the donor
and acceptor levels was analyzed in relation to time and analytical expressions were derived for the excitation transfer rate at a constant resonance
detuning and at a linear-in-time variation of the Stark shift of levels with consideration for the relaxation of the states via emission of phonons and photons.
In \cite{ref40} the Foster energy transfer process is investigated with consideration for the nonradiative and radiative contributions in the context
of the first order perturbation theory, where the electrostatic dipole-dipole coupling between the donor and the acceptor as well as their
electromagnetic interaction is taken into account.   It is demonstrated that for typical parameters of semiconductor quantum dots
the efficiency of the nonradiative transfer is close to the unity at the small donor-acceptor separation and for the energy detuning not larger
than 1-2 meV and decreases rapidly with an increase in the separation and the energy detuning. The radiative correction starts to play role only
at relatively long separations grater then 40 nm.

Forster-type energy transfer in assemblies of arrayed nanostructures is analyzed in the work \cite{add02}. Authors developed generalized theory for the nonradiative energy transfer in arrays with mixed dimensionalities.  The processes of energy transfer from a single
nanostructure (nanoparticle, nanowire, or quantum well) to 1D, 2D or 3D assemblies of nanoparticles and nanowire are studied paying particular attention to the functional
distance dependence of the transfer rate. It is shown that the dependence is determined by the quantum confinement as well as array stacking dimensionality of the acceptor. It is found that distance dependence of the transfer rate changes from $d^{-6}$ to $d^{-5}$ when the acceptor nanoparticles are arranged in a 1D stack, and to $d^{-4}$ when in a 2D array, and to $d^{-3}$ when in a 3D array, whereas the rate distance dependence for acceptor nanowires changes from $d^{-5}$ to $d^{-4}$ when they are arranged in a 1D stack, and to $d^{-3}$ when in a 2D array. 
It should be noted that the above-mentioned theoretical studies consider the resonance energy transfer due to direct Coulomb interaction
between quantum dots only and not taken into account their exchange interaction.

In paper \cite{add03} the mechanism of exciton tandem tunneling is proposed.  The corresponding rate for exciton hops as well as the Dexter (exchange) rate and the Forster one are evaluated for epitaxially connected nanocrystals. Evaluations carried out give the exciton hop rate larger than the Dexter rate and for Si even larger than the Forster rate.

In the present communication, we suggest a microscopic theory of the mechanism by which the nonradiative resonance energy transfer occurs between
spherical QDs based on a III--V semiconductor. Both the direct Coulomb and the exchange energy transfer mechanisms are considered.
To our knowledge, the detailed theoretical study of the nonradiative energy transfer between semiconductor quantum dots by the exchange mechanism is not presented
in the literature. The analytical treatment and numerical calculations of the exchange energy transfer rate for quantum dot system are first
performed in our study. We used the density-matrix method that enabled us to carry out an energy transfer analysis both in the weak-interaction
approximation, where the first order perturbation theory may be applied, and in the strong-interaction approximation.
The previously
employed models give no way of taking into account the real spectrum of III--V semiconductors and a number of the related new effects. We use the
Kane model as the model the most adequately describing the real spectrum of III--V semiconductors \cite{ref41,ref42}. We found the selection rules determining
the dipole-allowed and dipole-forbidden transitions in the acceptor and the dipole-allowed transition in the donor at which the energy-transfer matrix
element is nonzero. It is shown that taking into account the real band structure of the semiconductors extends the class of dipole-allowed and dipole-forbidden
transitions that are active in the energy transfer. The rates of both the direct Coulomb and the exchange energy transfer processes are calculated
by using the density-matrix formalism. For all the contributions, the dependence of the transfer rate on the distance between the energy donor and
acceptor are found. It is shown that at small distances the contribution of the exchange interaction between electrons in the donor and acceptor becomes important.

It is assumed that the donor QD and the acceptor QD are fabricated from the same III--V semiconductor compound and are
embedded in the matrix of another material that creates finite-depth potential wells for electrons ($V_{cD}$ ,  $V_{cA}$) and holes ($V_{hD}$ , $V_{hA}$).
The subscripts D and A correspond to the donor and acceptor. The system of two QDs under consideration is shown in Fig. 1. Figures 2 and 3 show schematically the processes
of the direct Coulomb and the exchange energy transfer, respectively. In the initial state of the system, the electron in the donor is in the excited
state (in the conduction band), and the electron in the acceptor is in the ground state (in the valence band). As a result of both the direct Coulomb
and the exchange interaction of the QDs, the excitation energy of the donor is transferred to the acceptor, and the system comes into the final state
in which the electron in the donor is in the valence band and the electron in the acceptor is in the conduction band. Also noteworthy is that the
nonradiative energy transfer process is similar to the Auger recombination in a QD, we considered previously in \cite{ref43}. However, in contrast to the
Auger recombination in which the interacting electrons are localized within a QD, the energy transfer occurs as a result of interaction between
electrons localized in different QDs, energy donor and acceptor.

\begin{figure}[h!]
\includegraphics[width=9cm]{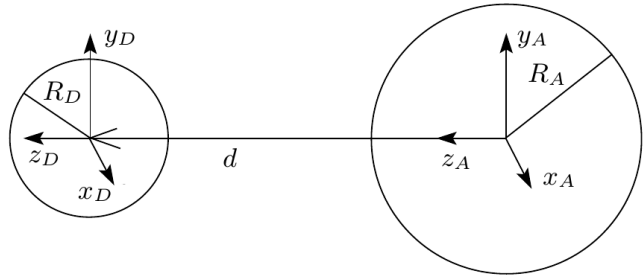}
\caption{Diagram of two QDs: donor of radius $R_D$ and acceptor of radius $R_A$.}
\end{figure}
\begin{figure}[h!]
\includegraphics[width=9cm]{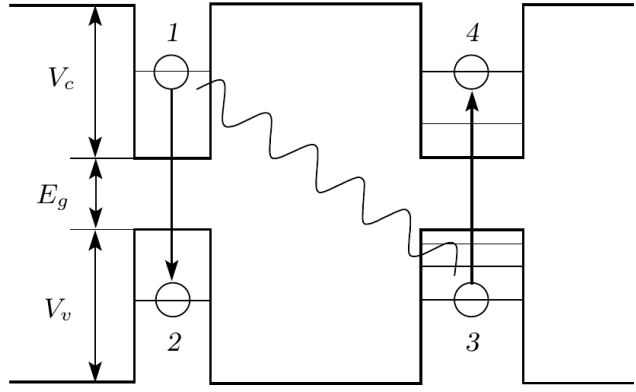}
\caption{Schematic of the energy transfer process in a two-quantum-dot system due to the direct Coulomb interaction}
\end{figure}
\begin{figure}[h!]
\includegraphics[width=8cm]{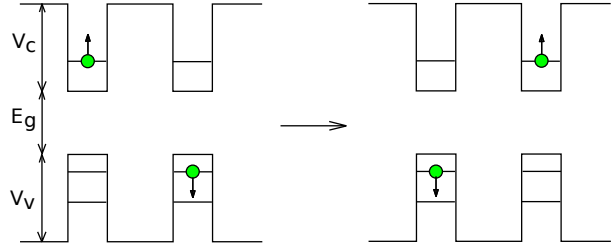}
\caption{Schematic of the energy transfer process in a two-quantum-dot system due to the exchange interaction}
\end{figure}

\section{Theory}

\subsection{Energy transfer matrix element}
To determine the rate of the nonradiative energy transfer between two QDS, it is necessary to calculate the energy transfer matrix element
(i.e., the matrix element of the Coulomb interaction) for the transition of the system from the initial to the final state (see Fig. 2). It can be expressed as
\begin{equation}
M_{if}=\sum\limits_{\bsigma_1,\bsigma_2}\int d^3r_1 d^3r_2 \psi_f^{*}(\xi_1,\xi_2)\frac{e^2}{\varepsilon |{\bf d+r_1-r_2}|}\psi_i(\xi_1,\xi_2),
\end{equation}
where $\xi_{i}=({\bf r_i},\sigma_i)$, ${\bf r_1}$ and ${\bf r_2}$ are the radius-vectors of electrons in the donor and acceptor, respectively, measured from the center of the corresponding QD;
$\bsigma_i$ are spin variables; $\varepsilon$ is the static dielectric constant of the medium. The antisymmetrized wave functions of the initial and final states
of the system under consideration are given by the expressions
\begin{equation}
\begin{aligned}
&\psi_i(\xi_1,\xi_2)=\frac{1}{\sqrt{2}}\left(\psi_{cD}({\bf r_1})\chi_{cD}(\bsigma_1)\psi_{hA}({\bf r_2})\chi_{hA}(\bsigma_2)\phantom{\frac{1}{1}}-\right.\\
&\left.\phantom{\frac{1}{1}}-\psi_{cD}({\bf r_2})\chi_{cD}(\bsigma_2)\psi_{hA}({\bf r_1})\chi_{hA}(\bsigma_1)\right),\\
&\psi_f(\xi_1,\xi_2)=\frac{1}{\sqrt{2}}\left(\psi_{hD}({\bf r_1})\chi_{hD}(\bsigma_1)\psi_{cA}({\bf r_2})\chi_{cA}(\bsigma_2)\phantom{\frac{1}{1}}-\right.\\
&\left.\phantom{\frac{1}{1}}-\psi_{hD}({\bf r_2})\chi_{hD}(\bsigma_2)\psi_{cA}({\bf r_1})\chi_{cA}(\bsigma_1)\right),
\end{aligned}
\end{equation}
Here, $\psi_{cD}({\bf r_i})$ and $\psi_{hD}({\bf r_i})$ are the wave functions of the spatial coordinates for electrons and holes in the donor (the functions for the acceptor can be written
similarly), and $\chi(\sigma_i)$ are spin wave functions. Substitution of wave functions (2) into the expression for matrix element (1) of the Coulomb interaction
results in its separation into two terms:
\begin{equation}
M_{if}=M_{coul}-M_{ex}
\end{equation}
where $M_{coul}$  is the matrix of the direct Coulomb interaction, and $M_{ex}$ is the matrix element of the exchange interaction. These matrix elements have the form
\begin{equation}\label{mcul1}
\begin{aligned}
&M_{coul}=\int d^3r_1 d^3r_2 \psi_{cD}({\bf r_1})\psi_{hD}^{*}({\bf r_1})\frac{e^2}{\varepsilon |{\bf d+r_1-r_2}|}\times\\
&\times\psi_{cA}^{*}({\bf r_2})\psi_{hA}({\bf r_2})\sum\limits_{\bsigma_1,\bsigma_2}\chi_{hD}^*(\bsigma_1)\chi_{cA}^*(\bsigma_2)\chi_{cD}(\bsigma_1)\chi_{hA}(\bsigma_2),
\end{aligned}
\end{equation}
and
\begin{equation}\label{mmmm}
\begin{aligned}
&M_{ex}=\int d^3r_1 d^3r_2 \psi_{cD}({\bf r_1})\psi_{cA}^{*}({\bf r_1})\frac{e^2}{\varepsilon |{\bf d+r_1-r_2}|}\times\\
&\times\psi_{hD}^{*}({\bf r_2})\psi_{hA}({\bf r_2})\sum\limits_{\bsigma_2,\bsigma_1}\chi_{hD}^*(\bsigma_2)\chi_{cA}^*(\bsigma_1)\chi_{cD}(\bsigma_1)\chi_{hA}(\bsigma_2).
\end{aligned}
\end{equation}
It follows from (4) that $M_{coul}$ is nonzero only for the transitions in which all spins remain unchanged, i.e., $\chi_{cD}=\chi_{hD}$  and  $\chi_{cA}=\chi_{hA}$. Equation (5) determines other
selection rules: $M_{ex}$ is not zero if $\chi_{cD}=\chi_{cA}$  and  $\chi_{hD}=\chi_{hA}$ . However, $\chi_{c(D,A)}$ should not be equal to  $\chi_{h(D,A)}$ and,
consequently, the spin functions on both QDs may vary simultaneously.
Ignoring the spin functions  $\chi$ in (5), we can see that this integral represents the electrostatic interaction between two clouds of charges, $Q_c=e\psi_{cD}^*({\bf r_1})\psi_{cA}({\bf r_1})$
and $Q_c=e\psi_{hA}^*({\bf r_2})\psi_{hD}({\bf r_2})$.

\subsection{Carrier wave functions}
In this study, we consider the nonradiative resonance energy transfer between spherical QDs fabricated from III--V semiconductors.
The most adequate description of the energy spectrum and wave functions of III--V semiconductors is provided by the Kane model \cite{ref41,ref42}.
In this model, the electron and hole wave functions can be written as
\begin{equation}
\psi=\psi_s\left|s\right>+\bpsi\left|{\bf p}\right>
\end{equation}
where $\left|s\right>$ and $\left|{\bf p}\right>$ are the Bloch wave functions of s- an p-types. Functions of the s-type describe states in the conduction band, and p-type
functions describe states in the valence band. The functions $\psi_s$  and  $\bpsi$ are envelope functions. In the spherical approximation, the Kane equations
for the envelope functions have the form \cite{ref43}:
\begin{equation}
\begin{aligned}
&(E_g+\delta-E)\psi_s-i\hbar \gamma \nabla\bpsi=0,\\
&-E\bpsi-i\hbar\gamma\nabla\psi_s+\frac{\hbar^2}{2m}(\gamma_1+4\gamma_2)\nabla\left(\nabla\bpsi\right)-\\
&-\frac{\hbar^2}{2m}(\gamma_1-2\gamma_2)\nabla\times(\nabla\times \bpsi)+i\delta{\bsigma}\times \bpsi=0,
\end{aligned}
\end{equation}
where  $\delta=\frac{\Delta_{so}}{3}$  ($\Delta_{so}$  is the spin-orbit coupling constant), $\gamma$ is the Kane matrix element having the rate dimension and related to the matrix element of the
momentum operator between states of the conduction band and valence band \cite{ref44}, $\gamma_1$ and $\gamma_2$ are generalized Luttinger parameters, and $m$ is the
free electron mass. Below, we consider the case in which the spin-orbit coupling constant  $\Delta_{so}=0$.  Later, we discuss the effect of the spin-orbit coupling
on the energy transfer process. 
Solutions to the Kane equations were also obtained in \cite{ref43}. The envelope functions of electrons within a QD are represented by the expressions:
\begin{equation}\label{l03}
\begin{aligned}
&\psi_s=Aj_j(k_c r)Y_{jm}(\theta,\phi),\\
&{\bpsi}=-\frac{i \hbar \gamma}{E_c+E_g}Ak_c \left(\sqrt{\frac{j+1}{2j+1}}j_{j+1}(k_c r){\bf Y}^{j+1}_{jm}(\theta,\phi)+\right.\\
&+\left.\sqrt{\frac{j}{2j+1}}j_{j-1}(k_c r){\bf Y}^{j-1}_{jm}(\theta,\phi)\right).
\end{aligned}
\end{equation}

Here, $Y_{jm}(\theta,\phi)$ are spherical functions; ${\bf Y}_{jm}^{j+1}(\theta,\phi)$ and
${\bf Y}_{jm}^{j-1}(\theta,\phi)$ are vector spherical harmonics; $j$ and $m$ are values of the full angular momentum and its projections on the $z$ axis,
respectively; $j_j(k_cr)$ are spherical Bessel functions, where $k_c$ is the wave number for the electron, $E_g$ is the energy gap width; $E_c$ is the electron energy
reckoned from the conduction band bottom; and $A$ is a normalizing constant. The envelope functions for electrons under the barrier:
\begin{equation}
\begin{aligned}
&\psi_s=Bk_j(\kappa_c r)Y_{jm}(\theta,\phi)\\
&\bpsi=\frac{-i\hbar\gamma\nabla\psi_s}{\tilde{E}+\tilde{E_g}}\\
&\nabla\psi_s=B\kappa_c\left(-\sqrt{\frac{j+1}{2j+1}}k_{j+1}(\kappa_c r){\bf Y}_{jm}^{j+1}+\right.\\
&\left.+\sqrt{\frac{j}{2j+1}}k_{j-1}(\kappa_c r){\bf Y}_{jm}^{j-1}\right),\\
&\tilde{E_g}=E_g+V_c+V_v,\\
&\tilde{E_c}=E_c-V_c.
\end{aligned}
\end{equation}

Here, $k_j(\kappa_c r)$ is the modified spherical Bessel function, $\kappa_c$ is the pseudowave number, $B$ is a normalizing constant, and $\tilde{E_c}$ is the electron energy reckoned
from the conduction band bottom in the wide-bandgap region.

In the three-band Kane model, the states of heavy holes are doubly degenerate because the spin-split band merges with the band of heavy holes.
The corresponding wave functions $\psi_{h1}$ and $\psi_{h2}$ contain no $\psi_s$ component and have different polarizations, which are determined by the polarizations of the
vector spherical harmonics:
\begin{equation}
\begin{aligned}
&\bpsi_{h1}=A_1 j_j(k_h r){\bf Y}_{jm}^{j}(\theta,\phi),\\
&{\bpsi_{h2}}=A_2 \left(\sqrt{\frac{j}{2j+1}}j_{j+1}(k_h r){\bf Y}^{j+1}_{jm}(\theta,\phi)-\right.\\
&-\left.\sqrt{\frac{j+1}{2j+1}}j_{j-1}(k_h r){\bf Y}^{j-1}_{jm}(\theta,\phi)\right),
\end{aligned}
\end{equation}
where $k_h$ is the wave number of a hole, and $A_1$ and  $A_2$ are normalization constants. The wave functions of heavy holes under the barrier were found
in the form \cite{ref43}:
\begin{equation}
\begin{aligned}
&\bpsi_{h1}=B_1 k_j(\kappa_h r){\bf Y}_{jm}^{j}(\theta,\phi),\\
&{\bpsi_{h2}}=B_2 \left(\sqrt{\frac{j}{2j+1}}k_{j+1}(\kappa_h r){\bf Y}^{j+1}_{jm}(\theta,\phi)-\right.\\
&-\left.\sqrt{\frac{j+1}{2j+1}}k_{j-1}(\kappa_h r){\bf Y}^{j-1}_{jm}(\theta,\phi)\right).
\end{aligned}
\end{equation}
In the Kane model, the principal quantum numbers  $n_c$  and  $n_h$ are introduced as  $n$-th root of the dispersion relationship for electrons and holes, respectively.
The dispersion relationship has the form
\begin{equation}\label{eq05}
\begin{aligned}
&j_j(k_c R)\left[\kappa_c \tilde{Z}\left(\frac{j k_{j}(\kappa_c R)}{\kappa_c R}-k_{j+1}(\kappa_c R)\right)\right]=\\
&=k_j(\kappa_c R)\left[k_c Z\left(\frac{j j_{j}(k_c R)}{k R}-j_{j+1}(k_c R)\right)\right],
\end{aligned}
\end{equation}
Here, $Z=1/(\mathscr{E}+E_g)$  to the left of the barrier, $\tilde{Z}=1/(\mathscr{E}+E_g+V_v)$ to the right of the barrier, and  $\mathscr{E}$ is the electron energy reckoned from the conduction band bottom.
The wave number $k_c$ and the pseudo-wave number $\kappa_c$ are given by
\begin{equation}
k_c^2=\frac{\mathscr{E}(\mathscr{E}+E_g)}{\hbar^2\gamma^2}
\end{equation}
and 
\begin{equation}
\kappa_c^2=\frac{(V_c-\mathscr{E})(\mathscr{E}+E_g+V_v)}{\hbar^2\gamma^2}
\end{equation}
The dispersion relationship for holes has the form
\begin{equation}\label{eq06}
\begin{aligned}
&j_j(k_h R)\frac{\kappa_h}{k_h}\left[j\left( \frac{(j+1)k_{j+1}(\kappa_h R)}{\kappa_h R}-k_{j+2}(\kappa_h R)\right)\right.+\\
&\left.+(j+1)\left(\frac{(j-1)k_{j-1}(\kappa_h R)}{\kappa_h R}-k_j(\kappa_h R)\right)\right]=\\
&=k_j(\kappa_h R)\frac{k_h}{\kappa_h}\left[j\left( \frac{(j+1)j_{j+1}(k_h R)}{k_h R}-j_{j+2}(k_h R)\right)-\right.\\
&\left.-(j+1)\left(\frac{(j-1)j_{j-1}(k_h R)}{k_h R}-j_j(k_h R)\right)\right],
\end{aligned}
\end{equation}                                                                                                                              (21)
where the wave number $k_h$  and the pseudo-wave number  $\kappa_h$ for holes
\begin{equation}
k_h^2=-2\frac{m_h E_h}{\hbar^2},
\end{equation}
\begin{equation}
\kappa_h^2=\frac{2m_h(V_v-E_h)}{\hbar^2}.
\end{equation}
The quantum-confinement levels for electrons and holes are determined by equations (12) and (15), together with the dispersion laws for electrons and holes \cite{ref43}.

\subsection{Matrix element of direct Coulomb interaction}
Because the matrix element of the direct Coulomb interaction contains the Coulomb interaction operator dependent on the variables  ${\bf r_1}$  and ${\bf r_2}$,
it is convenient, in order to separate the integration over these variables, to use the representation of the Coulomb interaction via the Fourier integral:
\begin{equation}
\begin{aligned}
&\frac{e^2}{\varepsilon |{\bf d+r_1-r_2}|}=\\
&=\frac{1}{2\pi^2\varepsilon}\int d^3q \frac{1}{q^2}\exp(i {\bf q} \cdot ({\bf d+r_1-r_2})).
\end{aligned}
\end{equation}
Then, the matrix element takes the form
\begin{equation}
M_{coul}=\frac{1}{2\pi^2\varepsilon}\int d^3q \exp(i {\bf q}\cdot {\bf d})\frac{1}{q^2}I_D(q)I_A^*(q),
\end{equation}
where the overlapping integrals for the donor and acceptor take the following form
\begin{equation}
\begin{aligned}
&I_D=\int d^3r_1 \exp (i {\bf q}\cdot {\bf r_1})\psi_{cD}({\bf r_1})\psi_{hD}^*({\bf r_1}),\\
&I_A=\int d^3r_2 \exp (i {\bf q}\cdot {\bf r_2})\psi_{cA}({\bf r_2})\psi_{hA}^*({\bf r_2}).
\end{aligned}
\end{equation}
Our consideration is performed in the framework of the Kane model, and, therefore, we take into account the admixture of valence band and conduction band
states. The electron wave function contains both the s- and p-components [see (6)]; the heavy hole wave functions contain only the p-component.

To derive an expression for the integral $I_{D1}$  without consideration for the mixing-in of valence band and conduction band states, the following should be
performed. First, we use the long-wavelength approximation by assuming that  $q a \ll 1$ ($a$  is the lattice constant), which makes it possible to separate the
integration over the rapidly oscillating Bloch function and the slowly varying envelope function. Further, it can be noted that the first term in the
Taylor series expansion of the exponential gives no contribution to the overall integral because the conduction-band and valence-band functions are orthogonal.
Therefore, the contribution is made by the second term in the expansion, which takes the form
\begin{equation}\label{eqnewid}
\begin{aligned}
I_{D1}\approx \int d^3 r_1 \bpsi_{hD}^{*}({\bf r_1})\psi_{cSD}({\bf r_1})\exp(i {\bf q r_1})\left<{\bf p}\right|i {\bf q r_{\alpha}}\left|s\right>.
\end{aligned}
\end{equation}
The integral $I_{D1}$  is proportional to the matrix element of the coordinate operator, $\left<{\bf p}\right|{\bf r_{\alpha 1}}\left|s\right>$, which can be expressed through the parameters of the semiconductor \cite{ref44}
\begin{equation}
\left<s\right|z\left|Z\right>=\frac{P}{E_g}.
\end{equation}
Here, $Z$ is one of the Bloch functions of the valence band (the others are $X$, $Y$). These functions are transformed as the corresponding coordinates, and $P=\hbar \gamma$ is
the Kane parameter related to the matrix element of the momentum operator between the states of the conduction band and valence band. For simplicity and convenience
of further calculations, we considered the crystallographic axes in the QDs to be codirectional, which results in that the full matrix element is independent
of the mutual orientation of the dipole momenta. If the angular dependence and the averaging over angles are taken into account, a factor of 2/3 appears
in the matrix element. Therefore, the integral $I_{D1}$ can be written as
\begin{equation}\label{eqid1}
\begin{aligned}
I_{D1}=i\frac{P}{E_g} \int d^3 r_1\left({\bf q}\cdot \bpsi_{hD}^{*}({\bf r_1})\right) \psi_{cSD}({\bf r_1})\exp(i {\bf q r_1}).
\end{aligned}
\end{equation}
Because we assume in our model that  $\Delta_{so}=0$, heavy holes are doubly degenerate. By virtue of this circumstance, there appear two different matrix elements
corresponding to transitions involving heavy holes with various polarizations. Let us designate the matrix element corresponding to transitions of holes
with the first and second polarizations as $M_{coul}^{(1)}$ and  $M_{coul}^{(2)}$, respectively. Calculations for the matrix element    are made in section "Calculation of direct-Coulomb matrix element". Except for the matrix
elements for the direct Coulomb interaction involving a heavy hole with the first or second polarization, two additional contributions for each polarization
of heavy holes appear in the full matrix element for the direct Coulomb interaction. We name these contributions the mixed-in contributions. One of these
is the contribution associated with only the mixing-in, and the second is the cross contribution. These contributions appear because the overlapping integral
for the donor and acceptor [formula (20)] contains a full electron wave function in the Kane model,which is a superposition of the envelope function and
s-type Bloch function and also of the envelope (vector) function and p-type Bloch function. This representation results in that two terms appear in each
overlap integral, and just this circumstance leads to the appearance of the above contributions. A detailed calculation of these contributions was made
in a separate communication \cite{ref45}. The admixture contribution was considered and it was shown that the contribution from these matrix elements is 5--10
times smaller than the contribution disregarding the admixture. Let us now calculate the above matrix elements $M_{Coul}^{(1)}$ and $M_{Coul}^{(2)}$.
The selection rules for the matrix element $M_{Coul}^{(1)}$ (\ref{A1}) have the form:
\begin{equation}\label{selrule1}
\left\{
\begin{aligned}
&m_{cA}=m_{hA},\\
&m_{cD}=m_{hD},\\
&l_1+j_{cD}+j_{hD}-\text{even},\\
&l_2+j_{cA}+j_{hA}-\text{even},\\
&\left|l_1-j_{cD}\right|\le j_{hD} \le l_1+j_{cD},\\
&\left|l_2-j_{cA}\right|\le j_{hA} \le l_2+j_{cA}.
\end{aligned}\right.
\end{equation}
The selection rules (24) follow from the symmetry properties of the Klebsch--Gordan coefficients \cite{ref57} [section "Calculation of direct-Coulomb matrix element", (\ref{A1})]. In the case of $l_1=0$ and  $l_2=0$, we
obtain the following simplified expression for the matrix element
\begin{equation}\label{m1s}
\begin{aligned}
&M_{coul}^{(1)}=\frac{e^2}{\varepsilon d^3}\mathscr{B}_{DA}^{(1)}\int\limits_0^{R_D} r_1^2 dr_1\int\limits_0^{R_A}r_2^2dr_2  \times\\
&\times\left(j_{j_{cD}}(k_{cD}r_1)j_{j_{hD}}(k_{hD}r_1)\right)\left(j_{j_{cA}}(k_{cA}r_2)j_{j_{hA}}(k_{hA}r_2)\right)\times\\
&\times\sqrt{\frac{2j_{cD}+1}{2j_{hD}+1}}\sqrt{\frac{2j_{cA}+1}{2j_{hA}+1}}C_{j_{hD},m_{hD},1,0}^{j_{hD},m_{hD}}C_{j_{hA},m_{hA},1,0}^{j_{hA},m_{hA}}\times\\
&\times\delta_{j_{cD},j_{hD}}\delta_{m_{cD},m_{hD}}\delta_{j_{cA},j_{hA}}\delta_{m_{cA},m_{hA}},
\end{aligned}
\end{equation}
where
\begin{equation}
\mathscr{B}_{DA}^{(i)}=2\left(\frac{P}{E_g}\right)^2 A_{cD}A_{hiD}A_{cA}A_{hiA},\,i=1,2.
\end{equation}
It follows from the selection rules (24) for the matrix element that the matrix element is not zero at $l_1=l_2=0$ for transitions involving electrons and
holes with equal angular momenta $j_{cD}=j_{hD}$  and $j_{cA}=j_{hA}$. These transitions are dipole-allowed. It follows for this case from expression (25) that the energy-transfer
matrix element depends on the donor--acceptror distance as  $M_{Coul}^{(1)}\sim 1/d^3$. If $l_1+l_2$ is odd, the sums of the angular momenta $j_{cD}+j_{hD}$  and $j_{cA}+j_{hA}$
of the states involved in the dipole
transition must have the opposite parity. The minimum admissible values of $l_1$  and $l_2$  are, respectively, 0 and 1. In this case, the restrictions imposed
on the allowed angular momenta in (24) have the form $j_{cD}=j_{hD}$  and  $|1-j_{cA}|\le |1+j_{cA}|$. Accordingly, the contribution to the matrix element of the Coulomb interaction can only
be made in the acceptor by dipole-forbidden transitions. When the above conditions are satisfied, $M_{Coul}^{(1)}$ depends on the donor--acceptor distance as  $1/d^4$.
It is clear that, for the transition to be possible, it is necessary that the transition energies in the donor and acceptor, which depend on the QD
radii $R_D$  and $R_A$  should coincide. In the case of the identical radii of QDs having the same system of energy levels, it is clear that the resonance energy
transfer becomes possible only when all the quantum numbers are the same.
Let us now consider the matrix element for a heavy hole with the second polarization. This matrix element is also calculated in section "Calculation of direct-Coulomb matrix element" [formula (\ref{A12}) and onward].

The symmetry properties of the Klebsch--Gordan coefficients \cite{ref57} give selection rules for the matrix element $M_{coul}^{(2)}$ (\ref{A12}):
\begin{equation}\label{sr2}
\left\{
\begin{aligned}
&m_{cA}=m_{hA},\\
&m_{cD}=m_{hD},\\
&l_1+j_{cD}+j_{hD}-\text{odd},\\
&l_2+j_{cA}+j_{hA}-\text{odd},\\
&\left|l_1-j_{cD}\right|\le j_{hD}\pm 1 \le l_1+j_{cD},\\
&\left|l_2-j_{cA}\right|\le j_{hA}\pm 1 \le l_2+j_{cA},\\
&\text{also, }j_{hD},j_{hA}\ge 1.
\end{aligned}\right.
\end{equation}
If $l_1+l_2=0$ , expression (\ref{A12}) is simplified to take the form
\begin{equation}\label{mif2simply}
\begin{aligned}
&M_{coul}^{(2)}=\frac{e^2}{\varepsilon d^3} \mathscr{B}_{DA}^{(2)} \int\limits_0^{R_D}r_1^2 dr_1\int\limits_0^{R_A}r_2^2 dr_2\times\\
&\times\left(j_{j_{cD}}(k_{cD}r_1)\left(\sqrt{\frac{j_{hD}}{2 j_{hD}+1}}j_{j_{hD}+1}(k_{hD}r_1)C_{j_{hD}+1,m_{hD},1,0}^{j_{hD},m_{hD}}\right.\right.\times\\
&\times\delta_{j_{cD},j_{hD}+1}\delta_{m_{cD},m_{hD}}-\sqrt{\frac{j_{hD}+1}{2j_{hD}+1}}j_{j_{hD}-1}(k_{hD}r_1)\times\\
&\times\left.\left.C_{j_{hD}-1,m_{hD},1,0}^{j_{hD},m_{hD}}\delta_{j_{cD},j_{hD}-1}\delta_{m_{cD},m_{hD}}\phantom{\sqrt{\frac{1}{1}}}\right)\right)\times\\
&\times\left(j_{j_{cA}}(k_{cA}r_2)\left(\sqrt{\frac{j_{hA}}{2 j_{hA}+1}}j_{j_{hA}+1}(k_{hA}r_2)C_{j_{hA}+1,m_{hA},1,0}^{j_{hA},m_{hA}}\right.\right.\times\\
&\times\delta_{j_{cA},j_{hA}+1}\delta_{m_{cA},m_{hA}}-\sqrt{\frac{j_{hA}+1}{2j_{hA}+1}}j_{j_{hA}-1}(k_{hA}r_2)\times\\
&\times\left.\left.C_{j_{hA}-1,m_{hA},1,0}^{j_{hA},m_{hA}}\delta_{j_{cA},j_{hA}-1}\delta_{m_{cA},m_{hA}}\phantom{\sqrt{\frac{1}{1}}}\right)\right).
\end{aligned}
\end{equation}
It follows from the selection rules (27) that the matrix element is not zero at $l_1=l_2=0$  only for the dipole-forbidden transitions in the donor and acceptor,
when the conditions $j_{cD}=j_{hD}\pm 1$ and $j_{cA}=j_{hA}\pm 1$ are satisfied. For this case, the matrix element $M_{Coul}^{(2)}$  (\ref{A14})
depends on the donor--aceptor distance as $1/d^3$ . When $l_1+l_2$  is an odd
number with the minimum possible values $l_1=0$  and  $l_2=1$, the matrix element is nonzero for the dipole-forbidden transitions in the donor if the condition  $j_{cD}=j_{hD}\pm 1$ is
satisfied; the following constraints are imposed on the transitions in the acceptor: $j_{cA}+j_{hA}$ is an even number and  $|1-j_{cA}|\le j_{hA}\pm 1 \le 1+j_{cA}$. When the above conditions are satisfied,
the matrix element $M_{Coul}^{(2)}$  depends on distance  $d$ as $1/d^4$. As already noted, for the matrix element to be nonzero, the relative values of $R_D$  and $R_A$  should provide
the resonance conditions for the corresponding transitions. In the case of QDs with the same radii, the resonance is observed for transitions between
levels with coinciding angular momenta, for electrons and holes of both QDs.

\subsection{Matrix element of exchange interaction}
When analyzing the contribution of the exchange interaction to the energy transfer, we restricted our consideration to a calculation of the matrix
element $M_{ex}$ with electron wave function $\psi_s$ (8) and heavy hole wave function $\psi_{h1}$ (10). Thus, we obtained the lower-bound estimate for the contribution of
the exchange interaction to the energy transfer rate.

Considering only the coordinate-related part in (5) and designating the functions $\psi_s$  and $\psi_{h1}$ for the donor and acceptor as $\psi_{csD}$, $\psi_{csA}$ and $\bpsi_{hD}$, $\bpsi_{hA}$, respectively,
we can write the matrix element $M_{ex}$ as: 
\begin{equation}\label{mmex}
\begin{aligned}
&M_{ex}=\int d^3 r'_1\int d^3 r_2 \psi_{csD}({\bf r_1})\psi_{csA}^{*}({\bf r_{1}})\frac{e^2}{\varepsilon\vert{\bf d+r_1-r_2}\vert}\times\\
&\times\psi_{hD}^{*}({\bf r_2})\psi_{hA}({\bf r_2})
\end{aligned}
\end{equation}
It is noteworthy that the radius vector ${\bf r_1}$  is reckoned from the donor center in the wave function of the donor, and ${\bf r_2}$, from the acceptor
center in the wave function of the acceptor. To calculate the matrix element, we place the origin of coordinates at the center of the acceptor
QD. In this coordinate system, (29) can be rewritten as
\begin{equation}\label{mmnew}
\begin{aligned}
&M_{ex}=\int d^3 r'_1\int d^3 r_2 \psi_{csD}({\bf r'_1-d})\psi_{csA}^{*}({\bf r'_{1}})\frac{e^2}{\varepsilon\vert{\bf r'_1-r_2}\vert}\times\\
&\times\psi_{hD}^{*}({\bf r_2-d})\psi_{hA}({\bf r_2})
\end{aligned}
\end{equation}
To simplify the calculation, we assume that the values of the spherical functions and vector spherical harmonics are equal to their values at $\theta_1=0$, $\theta_2=0$, $\theta_1'=\pi$ and $\theta_2'=\pi$,
which are determined by the angular momenta J and their projections on the z axis \cite{ref57}:
\begin{equation}\label{l66}
\begin{aligned}
&Y_{jm}(0,\phi)=\delta_{m0}\sqrt{\frac{2j+1}{4\pi}},\\
&Y_{jm}(\pi,\phi)=(-1)^{j}\delta_{m0}\sqrt{\frac{2j+1}{4\pi}},\\
&{\bf Y}_{jm}^{j}(0,\phi)=\left\{
\begin{aligned}
&-m\sqrt{\frac{2j+1}{8\pi}}{\bf e_m},\text{if }m=\pm 1,\\
&0-\text{ in the other cases}.\\
\end{aligned}
\right.\\
&{\bf Y}_{jm}^j(\pi,\phi)=(-1)^j{\bf Y}_{jm}^j(0,\phi).
\end{aligned}
\end{equation}
It should be noted that this approximation enables calculations only for $m_{cD}=m_{cA}=0$ and  $m_{hD}=m_{hA}=\pm 1$.
At other projections of the angular momentum on the axis, the more precise  $Y_{jm}(\theta,\phi)$  and ${\bf Y}_{jm}^j(\theta,\phi)$ should be taken into account.

It is convenient to solve our problem in the cylindrical system of coordinates because the system of two QDs under consideration has an axis that connects the QD centers.
In the cylindrical system of coordinates, the z-dependent parts of the donor electron wave functions in the donor region and under the barrier can be written as
\begin{equation}
\begin{aligned}
&\phi_{csD}(d-z_1')=A_{cD}j_{j_{cD}}(k_{cD}(d-z_1')),\text{ } d-R_D\le z_1'\le d,\\
&\phi_{csD}(d-z_1')=B_{cD}k_{j_{cD}}(\kappa_{cD}(d-z_1')),\text{ } 0\le z_1'\le d-R_D.
\end{aligned}
\end{equation}
The boundary conditions for these functions have the form
\begin{equation}
A_{cD}j_{j_{cD}}(k_{cD}R_D)=B_{cD}k_{j_{cD}}(\kappa_{cD}R_D).
\end{equation}
The z-dependent parts of the acceptor electron wave functions, donor and acceptor hole wave functions, and the boundary conditions for these
functions can be written in a similar way. In the cylindrical system of coordinates, the Coulomb potential of the system under consideration
is represented by the expression
\begin{equation}
\frac{e^2}{\varepsilon r}=\frac{e^2}{\varepsilon \sqrt{p^2+z^2}}.
\end{equation}
Here, $r=\left|{\bf r_1'-r_2}\right|$, $z=z_1'-z_2$, $p^2=\rho_1^2+\rho_2^2-2\rho_1\rho_2 \cos(\varphi_1-\varphi_2)$.
Using the integral formula
\begin{equation}
\frac{1}{\sqrt{p^2+z^2}}=\int\limits_0^\infty e^{-q |z|}J_0(q p)dq,
\end{equation} 
where $J_0(q p)$ is the zero-order Bessel function, we can rewrite the matrix element as
\begin{equation}\label{eqB}
\begin{aligned}
&M_{ex}\approx \frac{e^2}{\varepsilon}S\int \rho_1 d\rho_1 d\varphi_1dz'_1\int \rho_2 d\rho_2d\varphi_2dz_2\times\\
&\times\int\limits_0^\infty dq \phi_{cD}(d-z'_1)\phi_{csA}^{*}(z'_1)\times\\
&\times\exp(-q |z'_1-z_2|)J_0(q p)\phi_{hD}^{*}(d-z_2)\phi_{hA}(z_2).
\end{aligned}
\end{equation}
The quantity S is represented by the expression
\begin{equation}\label{sr1}
\begin{aligned}
&S=(-1)^{j_{cD}+j_{hD}+m_{hD}+1}\delta_{m_{cD},0}\sqrt{\frac{2j_{cD}+1}{4\pi}}\times\\
&\times\delta_{m_{cA},0}\sqrt{\frac{2j_{cD}+1}{4\pi}}\times\\
&\times(\delta_{m_{hD},1}+\delta_{m_{hD},-1})\sqrt{\frac{2j_{hD}+1}{8\pi}}\sqrt{\frac{2j_{hA}+1}{8\pi}}\delta_{m_{hD},m_{hA}}.
\end{aligned}
\end{equation}
\begin{figure}[t!]
\includegraphics[width=8cm]{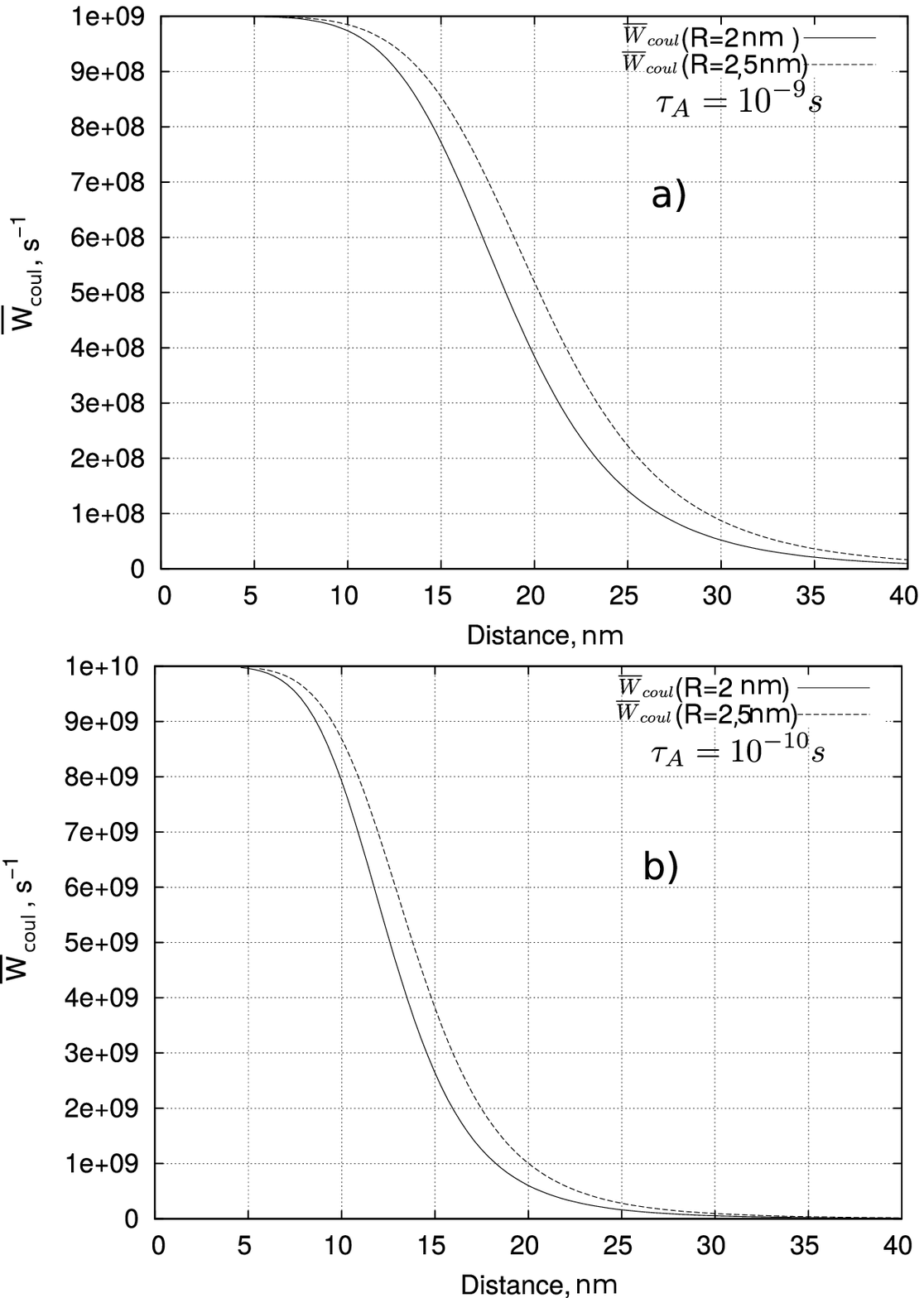}
\caption{Dependence of the rate of direct Coulomb energy transfer on the distance between the QDs.
The calculation was made for donor and acceptor transitions with quantum numbers $n_c=n_h=1$ for the wave function of heavy holes with the  second polarization, $\psi_{h2}$, for the ground transition.
The following radii were used $R_D=R_A=2,\,2.5$ nm. $M_{coul}^{(2)}$ was calculated for the values of the angular momentum and its projection (0,0) and (1,0).
The transverse relaxation time was taken to be $T_2=10^{-11}$ s.}
\end{figure}

\begin{figure}[t!]
\includegraphics[width=8cm]{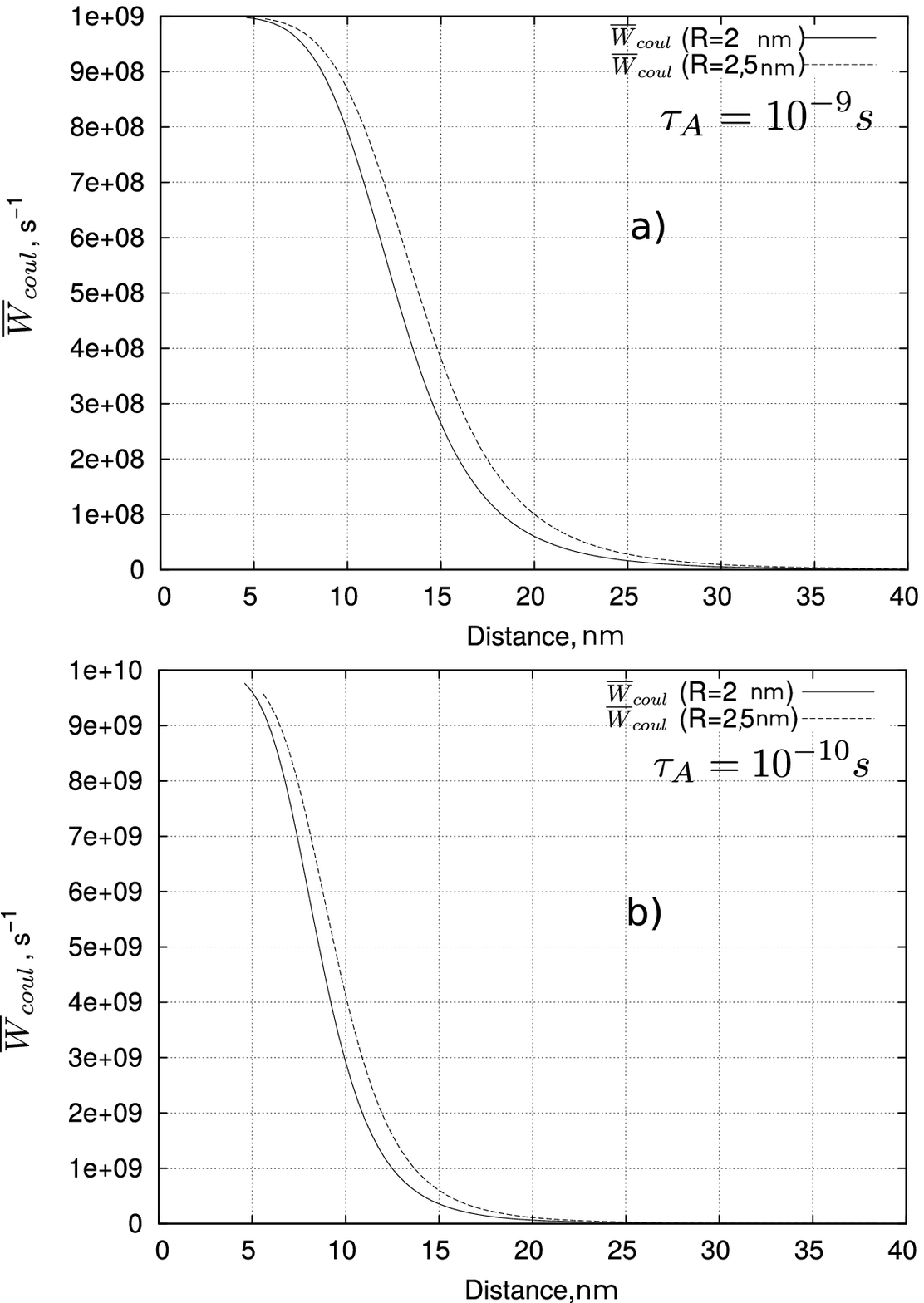}
\caption{Dependence of the rate of direct Coulomb energy transfer on the distance between the QDs.
The calculation was made for donor and acceptor transitions with quantum numbers $n_c=n_h=1$ for the wave function of heavy holes with the  second polarization, $\psi_{h2}$, for the ground transition.
The following radii were used $R_D=R_A=2,\,2.5$ nm. $M_{coul}^{(2)}$ was calculated for the values of the angular momentum and its projection (0,0) and (1,0).
The transverse relaxation time was taken to be $T_2=10^{-12}$ s.}
\end{figure}
In accordance with Graf's summation theorem \cite{ref46}, $J_0(q p)$  can be represented as
\begin{equation}\label{eqA}
J_0(q p)=\sum\limits_{n=-\infty}^{\infty}J_n(q \rho_1)J_n(q \rho_2)e^{i n (\varphi_1-\varphi_2)},
\end{equation}
where the variables $\rho_1$, $\rho_2$, $\varphi_1$, $\varphi_2$ are separated. Substituting this expression in (36), we bring the matrix element to the form
\begin{equation}\label{eq64}
\begin{aligned}
&M_{ex}\approx \frac{e^2}{\varepsilon}4\pi^2 S\int \rho_1 d\rho_1 dz'_1\int \rho_2 d \rho_2 dz_2\times\\
&\times\int\limits_0^\infty dq \phi_{csD}(d-z'_1)\phi_{csA}^{*}(z'_1)\exp(-q |z'_1-z_2|)\times\\
&\times J_0(q \rho_1)J_0(q \rho_2)\phi_{hD}^{*}(d-z_2)\phi_{hA}(z_2),
\end{aligned}
\end{equation}
because all the integrals over $\varphi_1$   and $\varphi_2$ are zero for all  $n$ except  $n=0$.
The integration range over $\rho$ is determined by the approximation chosen for the functions $Y_{jm}(\theta,\phi)$
and ${\bf Y}_{jm}^j (\theta,\phi)$ , i.e., by the condition $\rho_{max}\ll d$. To calculate the integrals take over $\rho$ in (39), it is useful to employ the following representation \cite{ref46}:
\begin{equation}
\int\limits_0^{\rho_{max}} \rho_{1(2)} J_0(q \rho_{1(2)})d\rho_{1(2)}=\frac{\rho_{max}}{q}J_1(q \rho_{max}).
\end{equation}
As a result, the matrix element can be obtained in the form
\begin{equation}\label{eqC}
\begin{aligned}
&M_{ex}\approx \frac{e^2}{\varepsilon}4\pi^2 S \rho_{max}^2\int\limits_0^\infty \frac{dq}{q^2}J_1^2(q \rho_{max})\times\\
&\times\int dz'_1 \int dz_2 \phi_{csD}(d-z'_1)\phi_{csA}^*(z'_1)\times\\
&\times e^{-q|z'_1-z_2|}\phi_{hD}^*(d-z_2)\phi_{hA}(z_2).
\end{aligned}
\end{equation}
Integration in (40) over $q$ yields \cite{ref47}:
\begin{equation}
\begin{aligned}
&\int\limits_0^\infty \frac{dq}{q^2}J_1^2(q \rho_{max})\exp(-q |z'_1-z_2|)=\\
&=\rho_{max}\left\{\frac{4}{3\pi}(\eta^2+1)^{1/2}\left[\eta^2 K\left((\eta^2+1)^{-1/2}\right)+\right.\right.\\
&+\left. \left. (1-\eta^2)E\left((\eta^2+1)^{-1/2}\right)\right]-\eta\right\}.
\end{aligned}
\end{equation}
Here, $\eta=(|z_1'-z_2|)/(2\rho_{max})$, $K(\xi)$ and $E(\xi)$ are full elliptical integrals of the first and second order. Let us introduce the following quantities
\begin{equation}
\begin{aligned}
&P(\eta_{1(2)}^2)=\rho_{max}\left\{\frac{4}{3\pi}(\eta_{1(2)}^2+1)^{1/2}\right.\times\\
&\times\left.\left[\eta_{1(2)}^2 K\left((\eta_{1(2)}^2+1)^{-1/2}\right)+(1-\eta_{1(2)}^2)E\left((\eta_{1(2)}^2+1)^{-1/2}\right)\right]\right\},
\end{aligned}
\end{equation}
where $\eta_1=(z_1'-z_2)/(2\rho_{max})$ for $z_1'>z_2$ and $\eta_2=(z_2-z_1')/(2\rho_{max})$ for $z_2>z_1'$.
Then, matrix element (41) can be rewritten as
\begin{equation}\label{l75}
\begin{aligned}
&M_{ex}\approx\frac{e^2}{\varepsilon}4\pi^2 S \rho_{max}^3\\
&\left[\int\limits_0^{R_A}\right.\left.\phi_{csD}^{*}(d-z'_1)\phi_{csA}(z'_1)dz'_1+\right.\int\limits_{R_A}^{d-R_D}\phi_{csD}(d-z'_1)\phi_{csA}^{*}(z'_1)dz'_1+\\
&+\left.\int\limits_{d-R_D}^{d}\phi_{csD}(d-z'_1)\phi_{csA}^{*}(z'_1)dz'_1\right]\cdot\left[\int\limits_0^{z'_1}\phi_{hD}^{*}(d-z_2)\right.\times\\
&\times\left.\phi_{hA}(z_2)(P(\eta_1^2)-\eta_1)dz_2\right.+\\
&+\left.\int\limits_{z'_1}^{d}\phi_{hD}^{*}(d-z_2)\phi_{hA}(z_2)(P(\eta_2^2)-\eta_2)dz_2\right]\equiv\\
&\equiv\frac{e^2}{\varepsilon}4\pi^2 S \rho_{max}^3[J_1+J_2+J_3]
\end{aligned}
\end{equation}
The $J_1$, $J_2$, and $J_3$ determine the contributions to the matrix element from the acceptor region, region between the QDs, and donor region, respectively.
In its turn the quantity $J_1$ can be written as 
\begin{equation}
\begin{aligned}
&J_1=\int\limits_0^{R_A}\phi_{csD}(d-z_1')\phi_{csA}^{*}(z_1')dz_1'\left[\int\limits_0^{z_1'}\phi_{hD}^{*}(d-z_2)\phi_{hA}(z_2)\right.\left.\left(P(\eta_1^2)-\eta_1\right)dz_2+\right.\\
&+\int\limits_{z_1'}^{R_A}\phi_{hD}^{*}(d-z_2)\phi_{hA}(z_2)\left(P(\eta_2^2)-\eta_2\right)dz_2+\\
&+\int\limits_{R_A}^d\left.\phi_{hD}^{*}(d-z_2)\phi_{hA}(z_2)\left(P(\eta_1^2)-\eta_1\right)dz_2\right].\\
\end{aligned}
\end{equation}
A detailed calculation of the integrals in (45) is made in section "Calculation of exchange integrals". The integrals appearing in J2, J3 are calculated in a similar manner. 
As a result of the cumbersome manipulations, the exchange interaction matrix element (44) is obtained in the following form:  
\begin{equation}
\begin{aligned}
&M_{ex}\approx\frac{e^2}{\varepsilon}2\pi^2S\rho_{max}^3\sin(k_{cD}R_D)\left(\frac{\sin(k_{hD}R_D)}{k_{hD}R_D}-\right.\\
&-\left.\cos(k_{hD}R_D)\right)\times\\
&\times\frac{1}{k_{cA}}\frac{1}{k_{cD}d}\frac{1}{k_{hA}}\frac{1}{k_{hD}d}\exp(-(\kappa_{cD}+\kappa_{hD})(d-R_D-R_A)).
\end{aligned}
\end{equation}
\subsection{Energy transfer probability}
Let us now consider the problem of calculating the rate of the nonradiative resonance energy transfer between QDs, i.e.,
the probability of transfer in unit time. We consider the energy transfer from a donor QD (D) to an acceptor QD (A). In what follows,
we are interested in the process of energy transfer in a system with irreversible energy transfer from the donor QD to the acceptor QD,
associated with the relaxation of excited states. To describe the energy transfer in a quantum-mechanical system with dissipation, it is
convenient to use the density-matrix method, which makes it possible to phenomenologically take into account both the relaxation processes
within the system and the interaction of the quantum-mechanical system with its environment \cite{ref01,ref48}. The equation for the density matrix $\widehat{\rho}$ has
in our case the following form \cite{ref01,ref49}:
\begin{equation}\label{l85}
\begin{aligned}
& i \hbar \frac{\partial\rho_{jj}}{\partial t}=[M_c,\rho]_{jj}+\frac{i\hbar}{T_1}(\rho_{jj}^{e}-\rho_{jj}),\\
& i \hbar \frac{\partial\rho_{ij}}{\partial t}=(E_i-E_j)\rho_{ij}+[M_c,\rho]_{ij}-\frac{i\hbar}{T_2}\rho_{ij}.
\end{aligned}
\end{equation}
Here, $M_c$ is the matrix element of the Coulomb interaction between the QDs, $\rho_{ii}$  are diagonal elements of the density matrix, $\rho_{ij}$  are its off-diagonal elements, $\rho_{jj}^{e}$
is the equilibrium value of the diagonal element of the density matrix, $T_1$  is the "longitudinal" relaxation time of the diagonal elements of the density
matrix (this the time of radiative and nonradiative transitions between levels, which determines the occupancy of states), $T_2$ is the "transverse" time
characterizing the relaxation of the off-diagonal elements of the density matrix, and $E_i-E_j$  is the energy difference between the initial and excited states.

Let us consider the density matrix for the following states (of the donor and acceptor):
\begin{equation}
\begin{aligned}
&\left|1\right>=\psi_D'\psi_A,\qquad\left|2\right>=\psi_D\psi_A',\qquad\left|3\right>=\psi_D\psi_A.\\
\end{aligned}
\end{equation}
Here, the prime is related to the excited states of the donor and acceptor. The  $\left|3\right>$  state in which both QDs are in the ground state (are not excited) is
necessary for retaining the normalization:  $\rho_{11}+\rho_{22}+\rho_{33}=1$. In our case (room temperature), the energy gap widths of the donor and acceptor QDs  $E_g^{D,A}\gg k_{B}T$($k_{B}$
is the Boltzmann constant, and  $T$ is absolute temperature). Then, it is apparent that the equilibrium values of the diagonal elements of the density matrix are given by  $\rho_{11}^e=\rho_{22}^e=0,\rho_{33}^e=1$.
As a result, we have from  (47) a system of equations for the elements of the density matrix \cite{ref01}.
\begin{figure}[t!]
\includegraphics[width=8cm]{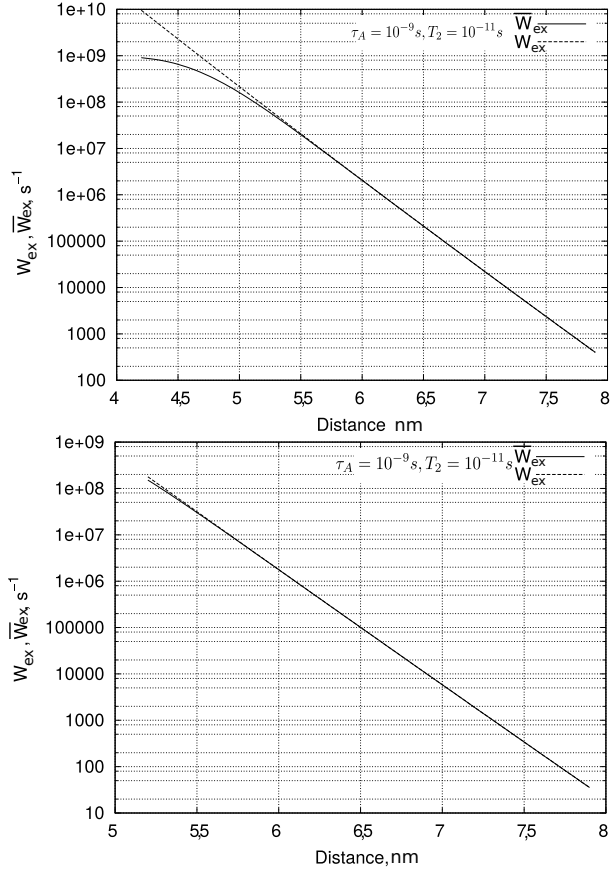}
\caption{Dependence of the rate of energy transfer by the exchange mechanism between two QDs.
The calculations were made for principal quantum numbers of electrons and holes $n_c=n_h=1$ and angular momenta $j_c=0$  and $j_h=1$.
The following radii were used: (a) $R_D=R_A=2$ nm, (b) $R_D=R_A=2.5$ nm. The radiative lifetime was taken to be $\tau_A=10^{-9}$ s, and the transverse relaxation time, to be  $T_2=10^{-11}$ s}
\end{figure}
\begin{figure}[t!]
\includegraphics[width=8cm]{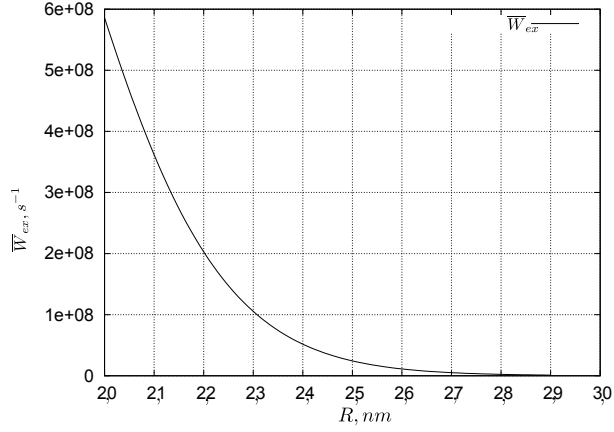}
\caption{Dependence of the rate of energy transfer by the exchange mechanism on the donor and acceptor radius at a near contact distance (($d = 2R+0.6$ nm) between the QDs.
The calculation was made for the principal quantum numbers of electrons and holes $n_c=n_h=1$  and angular momenta  $j_c=0$ and  $j_h=1$. The radiative lifetime was taken to be  $\tau_A=10^{-9}$ s,
and the transverse relaxation time, to be  $T_2=10^{-11}$ s.}
\end{figure}
\begin{equation}\label{ss}
\left\{
\begin{aligned}
&\frac{\partial \rho_{11}}{\partial t}=\frac{1}{i \hbar}((M_c)_{12}\rho_{21}-(M_c)_{21}\rho_{12})-\frac{\rho_{11}}{\tau_D},\\
&\frac{\partial \rho_{22}}{\partial t}=\frac{1}{i \hbar}((M_c)_{21}\rho_{12}-(M_c)_{12}\rho_{21})-\frac{\rho_{22}}{\tau_A},\\
&\frac{\partial \rho_{12}}{\partial t}=\frac{1}{i \hbar}(M_c)_{12}(\rho_{22}-\rho_{11})-\frac{\rho_{12}}{T_2}+\frac{\Delta E}{i \hbar}\rho_{12},\\
&\frac{\partial \rho_{21}}{\partial t}=\frac{1}{i \hbar}(M_c)_{21}(\rho_{11}-\rho_{22})-\frac{\rho_{21}}{T_2}-\frac{\Delta E}{i \hbar}\rho_{21}.
\end{aligned}
\right.
\end{equation}
Here, $\Delta E=E_{g}^{A}-E_g^{D}$ (resonance detuning), $\tau_D$  and $\tau_A$  are the longitudinal relaxation times for the donor and acceptor; the transverse relaxation time for two interacting QDs (D and A)
is related to halfwidths of levels in the donor and acceptor, $\Gamma_D$  and $\Gamma_A$, by $\frac{2}{T_2}=\frac{\Gamma_D}{\hbar}+\frac{\Gamma_A}{\hbar}$.
Let us analyze the system of equations (47) following \cite{ref01} . First, the general solution to
system  (49) has the nature of damped oscillations. Second, an excited donor can discharge energy (i.e., relax with a characteristic time  $\tau_l$) via two processes:
radiative recombination  $1/\tau_D$, or energy transfer to the acceptor, i.e.:
\begin{equation}\label{ww}
\frac{1}{\tau_l}\equiv \left[\int\limits_0^{\infty}\rho_{11}(t)dt\right]^{-1}=\frac{1}{\tau_D}+\overline{W},
\end{equation}
where $\overline{W}$ can be regarded as a generalized probability of energy transfer from the donor to the acceptor. Further, let us solve the system of equations (50) by the
method suggested in \cite{ref50}; for this purpose, we perform the Laplace transform of the density-matrix components:
\begin{equation}\label{laplace}
\begin{aligned}
&f_{ij}(s)=\mathscr{L}(\rho_{ij})=\int\limits_0^\infty \exp(-st)\rho_{ij}(t)dt,\\
&\int\limits_0^\infty\exp(-st)\frac{\partial\rho_{ij}}{\partial t} dt=sf_{ij}(s)-\rho_{ij}(0).
\end{aligned}
\end{equation}
As a result of the Laplace transform, the system of equations  (50) becomes
\begin{equation}\label{ssnew}
\left\{
\begin{aligned}
&sf_{11}-\rho_{11}(0)=\frac{1}{i \hbar}((M_c)_{12}f_{21}-(M_c)_{21}f_{12})-\frac{f_{11}}{\tau_D},\\
&sf_{22}-\rho_{22}(0)=\frac{1}{i \hbar}((M_c)_{21}f_{12}-(M_c)_{12}f_{21})-\frac{f_{22}}{\tau_A},\\
&sf_{12}-\rho_{12}(0)=\frac{1}{i \hbar}(M_c)_{12}(f_{22}-f_{11})-\frac{f_{12}}{T_2}+\frac{\Delta E}{i \hbar}f_{12},\\
&sf_{21}-\rho_{21}(0)=\frac{1}{i \hbar}(M_c)_{21}(f_{11}-f_{22})-\frac{f_{21}}{T_2}-\frac{\Delta E}{i \hbar}f_{21}.
\end{aligned}
\right.
\end{equation}
It should be noted that, according to (50) and (51), the generalized probability can be represented as
\begin{equation}
\overline{W}=-\frac{1}{\tau_D}+f_{11}^{-1}(0),
\end{equation}
where
\begin{equation}
f_{11}(0)=\int\limits_0^\infty \rho_{11}(t)dt.
\end{equation}
Assuming that s = 0 in  (52) and taking into account the initial conditions $\rho_{11}(0)=1$  and  $\rho_{ij}(0)=0$ at $i \neq 1$ or $j \neq 1$ , we obtain the following solution:
\begin{equation}
\begin{aligned}
&f_{11}^{-1}(0)=\left[\int\limits_0^\infty \rho_{11}(t)dt\right]^{-1}=\\
&=\frac{1}{\tau_D}+\frac{2|(M_c)_{12}|^2 T_2/\hbar^2}{1+\left(\frac{T_2\Delta E}{\hbar}\right)^2+\frac{2|(M_c)_{12}|^2}{\hbar^2}T_2 \tau_A}.
\end{aligned}
\end{equation}
As a result, we have for the generalized transfer probability $\overline{W}$:
\begin{equation}\label{wgeneral}
\overline{W}=\frac{2|(M_c)_{12}|^2 T_2/\hbar^2}{1+\left(\frac{T_2\Delta E}{\hbar}\right)^2+\frac{2|(M_c)_{12}|^2}{\hbar^2}T_2 \tau_A}.
\end{equation}
To analyze this solution and determine its physical meaning, let us consider a number of particular cases.

The first case corresponds to a small distance between the QDs, when their interaction is strong, so that
\begin{equation}
|M_c|\gg \frac{\hbar}{\tau_D},\frac{\hbar}{\tau_A},\frac{\hbar}{T_2}.
\end{equation}
In this case, solution  (49) for $\rho_{11}$  (at $\Delta E=0$) oscillates with frequency $\Omega=\frac{2 |M_c|}{\hbar}$; the energy transfer from the donor to the acceptor and back occurs in the system.
In this limiting case, the generalized probability $\overline{W}$ tends to  $1/\tau_A$; i.e., the energy transfer rate is determined by the rate at which the acceptor passes from the excited to the ground state.

At a weak interaction between the QDs, when  $\frac{2|M_c|^2}{\hbar^2}T_2\tau_A\ll 1$, we have, according to  (56):
\begin{equation}\label{wperturb}
\overline{W}=W=\frac{2|(M_c)_{12}|^2 T_2/\hbar^2}{1+\left(\frac{T_2\Delta E}{\hbar}\right)^2}.
\end{equation}
In this case, the energy transfer from a donor to an acceptor is an irreversible process and $W$ corresponds to the true probability of energy transfer in
unit time. It should be noted that expression  (58) can be derived in the framework of the ordinary perturbation theory. Expression  (58) can be rewritten as
\begin{equation}
W=\frac{2\pi}{\hbar}|(M_c)_{12}|^2\rho_f,
\end{equation}
where
\begin{equation}
\rho_f=\frac{1}{\pi}\frac{T_2/\hbar}{1+\frac{(T_2\Delta E)^2}{\hbar^2}}.
\end{equation}
Here,  $\rho_f$ has the meaning of the density of final states. Above, we calculated the matrix elements of the Coulomb interaction, which correspond to two polarizations of
heavy holes [$M_{coul}^{(1)}$ (\ref{A1}) and $M_{coul}^{(2)}$ (\ref{A12})] and the exchange matrix element $M_{ex}$ (44). The transitions involved in the energy transfer are determined by the corresponding selection
rules. Because all these contributions are independent, the full probability of energy transfer from a donor QD to an acceptor QD is given by
\begin{equation}\label{eq87}
W_{D\to A}=\sum\limits_{\alpha}W_{D\to A}^{\alpha},
\end{equation}
where $\alpha$ enumerates independent processes, and $W_{D \to A}^{\alpha}$ corresponds to the transfer probability for each of these.

\section{Results and discussion}
\subsection{Direct Coulomb energy transfer}
When the direct Coulomb interaction is considered, the matrix element for a heavy hole with the second polarization, $M_{coul}^{(2)}$ , is taken into account because the
transfer involving the ground state of the donor and acceptor for electrons and holes occurs just for this matrix element.
The calculations were made for the donor and acceptor based on the InAs material in a GaAs matrix, with radii $R_D=R_A=2$ $nm$ and $R_D=R_A=2.5$ $nm$ for
$V_c=V_v=0.52$ $eV$, $E_g=0.38$ $eV$, $m_e=0.03m_0$, $m_h=0.5m_0$. The calculation was performed
for two acceptor relaxation times $\tau_A=10^{-9}$ s and $\tau_A=10^{-10}$ s. The transverse relaxation time was taken to be $T_2=10^{-11}$ s (Fig.  4) and  $T_2=10^{-12}$ s
(Fig. 5). It should be noted that the
electron lifetime in the ground state ($10^{-9}$ s) is in good agreement with the time obtained in \cite{ref51,ref52}. The experimental values of this time have been repeatedly
reported. The authors of \cite{ref53} obtained a lifetime $\tau=882$ ps. This value is very close to the radiative time $\tau=0.7$  ns obtained in \cite{ref54}. The transverse relaxation time $T_2=10^{-11}$ s
is close to the value obtained in \cite{ref55,ref56}. Nevertheless, we allow variations of these two times by assuming the possible differences in the structure of both
the QD material itself and the embedding matrix.

The dependence of the energy transfer rate on the distance between the QDs, calculated by the formula for the generalized transfer probability   (58),
is shown in Figs. 4 and 5 for two values of $\tau_A$.  It can be seen in the figures that the longer the acceptor lifetime, the larger the distance between the QDs
at which the generalized probability $\overline{W_{coul}}$ approaches a constant value (saturates). It can also be seen in Figs. 4 and 5 that, at large distances between the QDs,
the generalized probability can be described by a formula of the perturbation theory, and hence follows that the probability of the Coulomb transfer obeys
the $1/d^6$ law at large distances, as follows from the Forster theory. The transfer rate at contact distances is determined as the inverse lifetime in the acceptor.
Note that  the direct Coulomb energy transfer matrix elements are calculated based  on the Kane model neglecting spin-orbit interaction. It was shown in \cite{ref45}
that the inclusion of the spin-orbit interaction to the Kane model results in multiplication  of the direct Coulomb energy transfer rate by the function $F(\Delta_{SO}/E_g)$,
which changes weakly for any relation between $\Delta_{SO}$ and $E_g$, having a maximum value $F(\Delta_{SO}/E_g)=1$ and minimum value $F(\Delta_{SO}/E_g)=0.9$.

\subsection{Exchange energy transfer}
Numerical calculations of the exchange energy transfer rate were performed for InAs QDs in a GaAs matrix with the same radii $R_D=R_A$ of 2 and 2.5 nm.
The same parameters of the system were used as those in calculations of the energy transfer via a direct Coulomb interaction: $V_c=V_v=0.52$ $eV$, $E_g=0.38$ $eV$, $m_e=0.03m_0$, $m_h=0.5m_0$.
The transitions between the energy levels in the donor and acceptor with principal quantum numbers $n_c=n_h=1$, angular momenta $j_c=0$, $j_h=1$, and
their projections $m_c=0$, $m_h=1$ were considered.
The matrix element of the exchange energy transfer was calculated by formula (44). The rate of the exchange energy transfer was found both by formula (56)
for the generalized transfer probability $\overline{W}_{ex}$  and by formula (58) for the transfer probability $W_{ex}$ at a weak interaction between the donor
and acceptor. Two
electron lifetimes in the ground state of the conduction band were taken ($\tau_A=10^{-9}$ s and $\tau_A=10^{-10}$ s) and two values of the
transverse relaxation time were assumed ($T_2=10^{-11}$ s or $T_2=10^{-12}$ s).

Figure 6 shows how the exchange energy transfer rate depends on the distance d between the donor and acceptor for QDs with radii $R_D=R_A=2$ nm. Calculations were made for
the transverse relaxation time $T_2=10^{-11}$ s and electron lifetime in the ground state of the conduction band $\tau_A=10^{-9}$ s. It can be seen that, at nearly contact distances d,
there is a noticeable difference between the transfer rate found from expression   (56)  for $\overline{W}_{ex}$ and the transfer rate found from expression  (58) for  $W_{ex}$.
$\overline{W}_{ex}$ exhibits saturation and tends to the rate of acceptor transition from the excited state to the ground state ($1/\tau_A$). Consequently,
the exchange interaction cannot be
considered as weak in this case. At large distances d, the rates  $\overline{W}_{ex}$  and $W_{ex}$ coincide. This indicates that the system parameters taken in our calculation satisfy
the weak interaction condition  $2|M_{ex}|^2 T_2\tau_A/(\hbar^2)\ll 1$  and formula  (58) can be used to calculate the transfer rate. The dependence of the transfer rate at distances d exceding
the nearly contact distances exhibits an exponential behavior. Calculation shows that the difference between $\overline{W}_{ex}$  and $W_{ex}$ at short d is substantially smaller for
the electron lifetime $\tau_A=10^{-10}$ s.  

For QDs with $R_D=R_A=2.5$ $nm$, the dependences of $\overline{W}_{ex}$ and $W_{ex}$ on distance d were calculated at the same transverse relaxation time  $T_2=10^{-11}$ s
and the same electron lifetimes $\tau_A=10^{-9}$ s and $\tau_A=10^{-10}$  s.
As shown by Fig. 7, the dependences of $\overline{W}_{ex}$ and $W_{ex}$, obtained at $\tau_A=10^{-9}$ s, are exponential and nearly coincide in the whole range of distances d.
The dependences of $\overline{W}_{ex}$ and $W_{ex}$ exhibit the same behavior at $\tau_A=10^{-10}$ s.
The exponential dependence obtained is consistent with the theoretical consideration for the exchange energy transfer between impurity
atoms or ions in an insulating crystal under the assumption that the interaction is weak \cite{ref05}. We can also note that experimental data for
organic molecules is well described as the exchange energy transfer with the exponential dependence on the distance using Perrin approximation \cite{ref62}.

The same behavior of the dependences of $\overline{W}_{ex}$ and $W_{ex}$ was obtained on the assumption of a shorter transverse relaxation time $T_2=10^{-12}$ s
at both times $\tau_A$  under consideration for QDs with both $R_D=R_A=2$ $nm$ and  $R_D=R_A=2.5$ $nm$. 

Thus, in all the case under consideration, except the first case with $\tau_A=10^{-9}$ s and $T_2=10^{-11}$ s, the system parameters taken in the calculation satisfy the weak interaction condition
$2|M_{ex}|^2 T_2\tau_A/(\hbar^2)\ll 1$
in the whole range of d and formula (58) can be used to calculate the energy transfer rate.                

Comparison of Fig. 4 for the rate of the direct Coulomb transfer and Fig. 6a for the rate of the exchange transfer between QDs with $R_D=R_A=2$ $nm$ shows that at small distances d,
at which the rates of transfer by both mechanisms experience a saturation, the contributions of these two mechanisms to the total transfer rate are the same for the
parameters chosen to be $\tau_A=10^{-9}$ s and $T_2=10^{-11}$ s. The exchange transfer rate is an order of magnitude lower than the direct Coulomb transfer rate at $d = 5.2$ $nm$
and is two orders of
magnitude lower at $d = 5.7$ $nm$. Thus, the contribution of the exchange transfer to the total transfer rate should be taken into account in numerical calculations at nearly
contact distances between the QDs. As the calculation demonstrated, the rate of the exchange transfer between QDs with  $R_D=R_A=2.5$ $nm$
at the same $\tau_A=10^{-9}$ s and $T_2=10^{-11}$ s for the contact distance d
is nearly an order of magnitude lower than that of the direct Coulomb transfer.

The dependence of the exchange energy transfer rate on the QD radius is determined by the quantum-confinement effect. It is known that, as the QD size increases, the energies
of the ground and excited states of electrons and holes become lower. This leads to an increase in the localizing potential and, as a consequence, to a weaker penetration of
wave functions into the subbarrier region. As a result, the overlap integral of the wave functions of the donor and acceptor decreases and so does the energy transfer rate.
Figure 7 shows the dependence of the energy transfer rate on the donor and acceptor radius $R_D=R_A$, caculated for the minimum distance d. It can be seen that
an increase in the QD radius from $R_D=R_A=2.0$ $nm$ to $R_D=R_A=2.5$ $nm$ leads to a substantial decrease in the exchange transfer rate.
The result is in agreement with data \cite{add04}, where experiments revealed faster energy transfer rates in smaller QD donor sizes.

The energy-transfer analysis made here and the set of plots make it possible to optimize parameters of the QD system for observing the maximum effect.

\section{Conclusions}

A microscopic theory of the mechanisms of nonradiative energy transfer between semiconductor quantum dots (QDs) based on III--V compounds is developed
in the framework of the three-band Kane model. Analytical expressions are derived both for the direct Coulomb interaction contributions and for the exchange
interaction contributions to the energy transfer matrix element in a system of two  spherical QDs fabricated from the same III--V semiconductor material and
embedded in a matrix of another semiconductor material. According to the selection rules obtained in the study, the energy transfer process can involve both
dipole-allowed and dipole-forbidden transitions in the donor and acceptor. In order to find the energy transfer rate, the density-matrix method is used.
For the Coulomb interaction  a numerical calculations of the generalized energy transfer  rate  for the lowest energy transition in the donor and acceptor
demonstrate a saturation at the nearly contact distances between the donor and acceptor. This indicates that the Coulomb
interaction cannot be considered as weak in this case. At larger distances $d$  between the QDs, the rate of the direct Coulomb energy transfer obeys the $1/d^6$  law  in accordace with the Forster theory.  The exchange energy transfer between QDs is first analyzed in detail. A numerical calculations show that the exchange energy transfer rate saturates at the nearly contact distances d  between the donor and acceptor and becomes exponential as d increases. At the small distances, the contribution of the exchange interaction to the energy transfer rate may be of the same order of magnitude as the contribution of the direct Coulomb interaction  and these contributions may have the same value in the saturation range. Consequently, it is important  to take  into account both of these contributions
in a quantitative description of the energy trasfer between QDs.

Two recent papers \cite{add04,add05}
support our result that in the systems including QDs the exchange interaction can play a significant role in the energy transfer
at small donor-acceptor separation. It is experimentally shown in
\cite{add04}
that in the CdSe QD-Squaraine light harvesting assemblies, the exchange (Dexter) process is essential to the energy
transfer and dominates the dipole-based Forster one at the smaller QDs. In
\cite{add05}
the measured energy transfer rates in the close-packed blends of  CdSe/CdZnS core/shell QDs are found to be more than an
order of magnitude larger than the rate predicted by Forster theory,  which cannot be satisfactory explained by several
possible factors considered in the paper. We believe that exchange contribution to the energy transfer should be also
taken into account for the consideration the discrepancy.

Thus our work enhances understanding of the features both of the direct coulomb and of the exchange energy transfer between quantum dots at their small
separations. The results can be importance for the developing of the high-performance electronics based on thin films
of the density packed QDs. Our results are also essential for the investigation of the structure of biomolecules and
their complexes.
The nonradiative energy transfer between the energy donor and acceptor  conjugated with biomolecules is widely used in medical and biological  experiments \cite{ref23,ref24,ref25,ref26}. The application of semiconductor QDs as both a donor and an acceptor improved the potentiality of these experiments. The high sensitivity of the transfer rate to changes in the distance between the energy donor and acceptor makes it possible to detect the formation of antigen-antibody and enzyme-substrate complexes and the DNA hybridization and to study the structure and dynamics of biomolecules, where it is necessary to measure short distances within a molecule \cite{ref24,ref58,ref59,ref60}. Results of studies of this kind are highly important for diagnostics and therapy of a number of diseases, including those of oncological nature \cite{ref61}. The development of an adequete energy transfer theory taking into account both the direct Coulomb interaction between electrons of a donor and an acceptor and their exchange interaction is necessary for correct interpretation of experimental data.

\section{Calculation of direct-Coulomb matrix element}
To calculate the overlap integral $I_{D1}$   and then the matrix element, the following steps should be taken. In the first step, it is convenient to substitute the
representation of the wave vector $q$ and the hole wave function in terms of the cyclic coordinates and then calculate the scalar product. In the second step,
the plane wave is expanded in spherical functions. In the third step, it is necessary to perform integration over angular variables and substitute the results
in the matrix element. So, we can represent the matrix element as
\begin{equation}\label{A1}
\begin{aligned}
&M_{coul}^{(1)}=\frac{2}{\pi}\frac{e^2}{\varepsilon}\left(\frac{P}{E_g}\right)^2 A_{cD}A_{h1D}A_{cA}A_{h1A}\times\\
&\times\int\limits_0^{R_D} r_1^2 dr_1\int\limits_0^{R_A}r_2^2dr_2 \times\\
&\times\frac{1}{3}\sum\limits_{l_1,l_2=0}^{\infty}\left(j_{j_{cD}}(k_{cD} r_1)j_{j_{hD}}(k_{hD} r_1)\right)\times\\
&\times\left(j_{j_{cA}}(k_{cA} r_2)j_{j_{hA}}(k_{hA} r_2)\right)\times\\
&\times\sqrt{\frac{2j_{cD}+1}{2j_{hD}+1}}\sqrt{\frac{2j_{cA}+1}{2j_{hA}+1}}C_{j_{hD},m_{hD},1,0}^{j_{hD},m_{hD}}C_{j_{hA},m_{hA},1,0}^{j_{hA},m_{hA}}\times\\
&\times(2l_1+1)(2l_2+1)i^{l_1-l_2}C_{l_1,0,j_{cD},0}^{j_{hD},0}C_{l_1,0,j_{cD},m_{cD}}^{j_{hD},m_{hD}}\times\\
&\times C_{l_2,0,j_{cA},0}^{j_{hA},0}C_{l_2,0,j_{cA},m_{cA}}^{j_{hA},m_{hA}}I(q),
\end{aligned}
\end{equation}
where
\begin{equation}\label{A2}
\begin{aligned}
&I(q)=\int dq q^2 d\Omega_q j_{l_1}(q r_1)j_{l_2}(q r_2)\exp(i q d \cos(\theta_q))\times\\
&\times(Y_{10}(\Omega_q))^2.
\end{aligned}
\end{equation}
Here, $R_D$ and $R_A$ are, respectively, the radii of the donor and acceptor QDs; $A_{cD}$ and $A_{h1D}$ are the normalization constants for the donor electron and hole wave functions
(similarly for the acceptor); and $k_{cD}$  and $k_{hD}$ are the electron and hole wave numbers of the donor (similarly for the acceptor). It is noteworthy that integration only
over the QD region in the matrix element is sufficient because the heavy hole wave function rapidly decays into the barrier. To integrate over $q$ in (\ref{A1}), we are
going to need a series expansion of the plane wave:
\begin{equation}\label{A3}
\begin{aligned}
&\exp(i q d \cos(\theta_q))=\\
&=\sqrt{4 \pi}\sum\limits_{l=0}^{\infty}i^l j_l(q d)\sqrt{2l+1}Y_{l0}(\Omega_q).
\end{aligned}
\end{equation}
The product of two spherical functions is expressed as \cite{ref54}:
\begin{equation}\label{A4}
\begin{aligned}
&(Y_{10}(\Omega_q))^2=\\
&=\sum\limits_{LM}\sqrt{\frac{9}{4\pi(2 L+1)}}C_{1010}^{L0}C_{1010}^{LM}Y_{LM}(\Omega_q).
\end{aligned}
\end{equation}
By virtue of the symmetry properties of the Clebsch--Gordan coefficients, only the coefficients with $L = 0, 2$ are nonzero in (\ref{A4}). In this case, $l=L=0,2$ . Therefore, integral (\ref{A2}) can be represented as
\begin{equation}\label{A5}
I(q)=\frac{1}{d^3}(I_0-2I_2),
\end{equation}
where
\begin{equation}\label{A6}
I_l=\int\limits_0^{\infty}t^2 dt j_l(t)j_{l_1}(t \frac{r_1}{d})j_{l_2}(t \frac{r_2}{d});
\end{equation}
and  $l$  takes two values: 0 and 2. Integral (6) can be expressed through the hypergeometric Appel function $F_4(a,b;c,d;x,y)$.
\begin{equation}\label{A7}
\begin{aligned}
&I_l=\pi^{\frac{3}{2}}\frac{1}{2}\left(\frac{r_1}{d}\right)^{l_1}\left(\frac{r_2}{d}\right)^{l_2}\times\\
&\times\left[\frac{\Gamma(\frac{l_1+l_2+l+3}{2})}{\Gamma(l_1+\frac{3}{2})\Gamma(l_2+\frac{3}{2})\Gamma(\frac{l-(l_1+l_2)}{2})}\right.\times\\
&\times F_4\left(\frac{l_1+l_2-l+2}{2},\frac{l_1+l_2+l+3}{2};\right.\\
&;\left.\left.l_1+\frac{3}{2},l_2+\frac{3}{2};\left(\frac{r_1}{d}\right)^2,\left(\frac{r_2}{d}\right)^2\right)\right],
\end{aligned}
\end{equation}
where $\Gamma(x)$  is the gamma function. The calculated integral enters into expression (\ref{A1}) for the matrix element. It follows from the properties of the gamma function that
the matrix element $M_{Coul}^{(1)}$  is nonzero if one of the conditions 
\begin{equation}\label{A8}
\left\{
\begin{aligned}
&1.\;l_1=l_2=0,\\
&2.\;l_1+l_2-\text{ odd.}\\
\end{aligned}
\right.
\end{equation}
is satisfied.
Substituting (\ref{A5}) into (\ref{A1}), we obtain the matrix element
\begin{equation}\label{A9}
\begin{aligned}
&M_{coul}^{(1)}=\frac{2}{\pi}\frac{e^2}{\varepsilon d^3}\left(\frac{P}{E_g}\right)^2 A_{cD}A_{h1D}A_{cA}A_{h1A}\times\\
&\times\int\limits_0^{R_D} r_1^2 dr_1\int\limits_0^{R_A}r_2^2dr_2  \times\\
&\times\frac{1}{3}\sum\limits_{l_1,l_2=0}^{\infty}\left(j_{j_{cD}}(k_{cD}r_1)j_{j_{hD}}(k_{hD}r_1)\right)\times\\
&\times\left(j_{j_{cA}}(k_{cA}r_2)j_{j_{hA}}(k_{hA}r_2)\right)\times\\
&\times\sqrt{\frac{2j_{cD}+1}{2j_{hD}+1}}\sqrt{\frac{2j_{cA}+1}{2j_{hA}+1}}C_{j_{hD},m_{hD},1,0}^{j_{hD},m_{hD}}C_{j_{hA},m_{hA},1,0}^{j_{hA},m_{hA}}\times\\
&\times (2l_1+1)(2l_2+1)i^{l_1-l_2}C_{l_1,0,j_{cD},0}^{j_{hD},0}C_{l_1,0,j_{cD},m_{cD}}^{j_{hD},m_{hD}}\times\\
&\times C_{l_2,0,j_{cA},0}^{j_{hA},0}C_{l_2,0,j_{cA},m_{cA}}^{j_{hA},m_{hA}}\left(I_0-2I_2\right).
\end{aligned}
\end{equation}
When $l_1=l_2=0$, it can be shown that
\begin{equation}\label{A10}
\begin{aligned}
&I_0=0,\\
&I_2=\pi^{3/2}\frac{1}{2}\frac{\Gamma(5/2)}{\Gamma(3/2)\Gamma(3/2)\Gamma(1)}=\frac{3\pi}{2}.
\end{aligned}
\end{equation}
and, therefore, the matrix element takes the form
\begin{equation}\label{A11}
\begin{aligned}
&M_{coul}^{(1)}=2\frac{e^2}{\varepsilon d^3}\left(\frac{P}{E_g}\right)^2 A_{cD}A_{h1D}A_{cA}A_{h1A}\int\limits_0^{R_D} r_1^2 dr_1\int\limits_0^{R_A}r_2^2dr_2  \times\\
&\times\sum\limits_{l_1,l_2=0}^{\infty}\left(j_{j_{cD}}(k_{cD}r_1)j_{j_{hD}}(k_{hD}r_1)\right)\left(j_{j_{cA}}(k_{cA}r_2)j_{j_{hA}}(k_{hA}r_2)\right)\times\\
&\times\sqrt{\frac{2j_{cD}+1}{2j_{hD}+1}}\sqrt{\frac{2j_{cA}+1}{2j_{hA}+1}}C_{j_{hD},m_{hD},1,0}^{j_{hD},m_{hD}}C_{j_{hA},m_{hA},1,0}^{j_{hA},m_{hA}}\times\\
&\times (2l_1+1)(2l_2+1)i^{l_1-l_2}C_{l_1,0,j_{cD},0}^{j_{hD},0}C_{l_1,0,j_{cD},m_{cD}}^{j_{hD},m_{hD}}\times\\
&\times C_{l_2,0,j_{cA},0}^{j_{hA},0}C_{l_2,0,j_{cA},m_{cA}}^{j_{hA},m_{hA}}.
\end{aligned}
\end{equation}

Let us now consider the matrix element $M_{Coul}^{(2)}$  with the wave function $\psi_{h2}$  of a heavy hole with the second polarization. This matrix element can be
calculated similarly to the matrix element $M_{Coul}^{(1)}$  and has the form
\begin{equation}\label{A12}
\begin{aligned}
&M_{coul}^{(2)}=\frac{e^2}{\varepsilon d^3}\frac{2}{3 \pi}\left(\frac{P}{E_g}\right)^2 A_{cD}A_{h2D}A_{cA}A_{h2A}\times\\
&\times\int\limits_0^{R_D}\int\limits_0^{R_A} dr_1 r_1^2 dr_2 r_2^2 \sum\limits_{l_1,l_2=0}^{\infty}\left(j_{j_{cD}}(k_{cD}r_1)\right.\times\\
&\times\left.\left(\sqrt{\frac{j_{hD}}{2j_{hD}+1}}j_{j_{hD}+1}(k_{hD}r_1)\right.\right.\times\\
&\times C_{j_{hD}+1,m_{hD},1,0}^{j_{hD},m_{hD}}\sqrt{\frac{2j_{cD}+1}{2j_{hD}+3}}C_{l_1,0,j_{cD},0}^{j_{hD}+1,0}C_{l_1,0,j_{cD},m_{cD}}^{j_{hD}+1,m_{hD}}-\\
&-\sqrt{\frac{j_{hD}+1}{2j_{hD}+1}}j_{j_{hD}-1}(k_{hD}r_1)C_{j_{hD}-1,m_{hD},1,0}^{j_{hD},m_{hD}}\sqrt{\frac{2j_{cD}+1}{2j_{hD}-1}}\times\\
&\times\left.\left.C_{l_1,0,j_{cD},0}^{j_{hD}-1,0}C_{l_1,0,j_{cD},m_{cD}}^{j_{hD}-1,m_{hD}}\phantom{\frac{1}{1}}\right)\right)\times\\
&\times\left(j_{j_{cA}}(k_{cA}r_2)\left(\sqrt{\frac{j_{hA}}{2j_{hA}+1}}j_{j_{hA}+1}(k_{hA}r_2)\right.\right.\times\\
&\times C_{j_{hA}+1,m_{hA},1,0}^{j_{hA},m_{hA}}\sqrt{\frac{2j_{cA}+1}{2j_{hA}+3}}C_{l_2,0,j_{cA},0}^{j_{hA}+1,0}\times\\
&\times C_{l_2,0,j_{cA},m_{cA}}^{j_{hA}+1,m_{hA}}-\sqrt{\frac{j_{hA}+1}{2j_{hA}+1}}j_{j_{hA}-1}(k_{hA}r_2)C_{j_{hA}-1,m_{hA},1,0}^{j_{hA},m_{hA}}\times \\
&\times \sqrt{\frac{2j_{cA}+1}{2j_{hA}-1}}\times\\
&\times \left.\left.C_{l_2,0,j_{cA},0}^{j_{hA}-1,0}C_{l_2,0,j_{cA},m_{cA}}^{j_{hA}-1,m_{hA}}\phantom{\frac{1}{1}}\right)\right) \times\\
&\times i^{l_1-l_2}(2l_1+1)(2l_2+1)\left(I_0-2I_2\right),
\end{aligned}
\end{equation}
where $I_0$  and  $I_2$ are represented by expressions (\ref{A10}). It can be shown that the selection rules for $M_{Coul}^{(2)}$  are as follows
\begin{equation}\label{A13}
\left\{
\begin{aligned}
&l_1+l_2=0,\\
&l_1+l_2-\text{ odd}
\end{aligned}
\right.
\end{equation}
In the case of $l_1+l_2=0$, expression (\ref{A12}) is simplified to become
\begin{equation}\label{A14}
\begin{aligned}
&M_{coul}^{(2)}=\frac{e^2}{\varepsilon d^3} \left(\frac{P}{E_g}\right)^2 A_{cD}A_{h2D}A_{cA}A_{h2A} \int\limits_0^{R_D}r_1^2 dr_1\int\limits_0^{R_A}r_2^2 dr_2 2\times\\
&\times\left(j_{j_{cD}}(k_{cD}r_1)\left(\sqrt{\frac{j_{hD}}{2 j_{hD}+1}}j_{j_{hD}+1}(k_{hD}r_1)C_{j_{hD}+1,m_{hD},1,0}^{j_{hD},m_{hD}}\right.\right.\times\\
&\times\delta_{j_{cD},j_{hD}+1}\delta_{m_{cD},m_{hD}}-\sqrt{\frac{j_{hD}+1}{2j_{hD}+1}}j_{j_{hD}-1}(k_{hD}r_1)\left.\left.C_{j_{hD}-1,m_{hD},1,0}^{j_{hD},m_{hD}}\delta_{j_{cD},j_{hD}-1}\delta_{m_{cD},m_{hD}}\phantom{\sqrt{\frac{1}{1}}}\right)\right)\times\\
&\times\left(j_{j_{cA}}(k_{cA}r_2)\left(\sqrt{\frac{j_{hA}}{2 j_{hA}+1}}j_{j_{hA}+1}(k_{hA}r_2)C_{j_{hA}+1,m_{hA},1,0}^{j_{hA},m_{hA}}\right.\right.\delta_{j_{cA},j_{hA}+1}\delta_{m_{cA},m_{hA}}-\sqrt{\frac{j_{hA}+1}{2j_{hA}+1}}j_{j_{hA}-1}(k_{hA}r_2)\times\\
&\times\left.\left.C_{j_{hA}-1,m_{hA},1,0}^{j_{hA},m_{hA}}\delta_{j_{cA},j_{hA}-1}\delta_{m_{cA},m_{hA}}\phantom{\sqrt{\frac{1}{1}}}\right)\right).
\end{aligned}
\end{equation}
\section{Calculation of exchange interaction integrals}
The integrals in (45) are calculated for transitions between levels with full angular momenta  $j_{cD} = j_{cA} = 0$ and $j_{hD} = j_{hA} = 1$.
In the radial parts of wave functions (32), we pass to cylindrical functions of half-integer order.  Let us $J_1$ as $J_1 = J_{11} + J_{12} + J_{13}$, where
\begin{equation}\label{B1}
\begin{aligned}
&J_{11}=\sin(k_{cD}R_D)\exp(-\kappa_{cD})\int\limits_0^{R_A}\frac{1}{k_{cD}(d-z_1'1)}\frac{\sin(k_{cA}z_1')}{k_{cA}z_1'}\exp(\kappa_{cD}z_1')dz_1'\times\\
&\times\left\{\exp(-\kappa_{hD}(d-R_D))\right.\times\\
&\times\left\{\left(\frac{\sin(k_{hD}R_D)}{k_{hD}R_D}-\cos(k_{hD}R_D)\right)\int_0^{R_A}\frac{1}{k_{hD}(d-z_2)}\frac{1}{k_{hA}z_2}\right.\times\\
&\times\left.\left(\frac{\sin(k_{hA}z_2)}{k_{hA}z_2}-\cos(k_{hA}z_2)\right)\exp(\kappa_{hD}z_2)P(\eta_2^2)dz_2 \right.+\\
&+\left(\frac{\sin(k_{hD}R_D)}{k_{hD}R_D}-\cos(k_{hD}R_D)\right)\left(\frac{\sin(k_{hA}R_A)}{k_{hA}R_A}-\cos(k_{hA}R_A)\right)\exp(-\kappa_{hD}(d-R_D))\exp(\kappa_{hA}R_A)+\\
&+\int\limits_{R_A}^{d-R_D}\frac{1}{k_{hD}(d-z_2)}\frac{1}{k_{hA}z_2}\exp((\kappa_{hD}-\kappa_{hA})z_2)P(\eta_2^2)dz_2+\left(\frac{\sin(k_{hA}R_A)}{k_{hA}R_A}-\cos(k_{hA}R_A)\right)\times\\
&\times\exp(k_{hA}R_A)\int\limits_{d-R_D}^{d}\frac{1}{k_{hA}z_2}\frac{1}{k_{hD}(d-z_2)}\left(\frac{\sin(k_{hD}(d-z_2))}{k_{hD}(d-z_2)}-\cos(k_{hD}(d-z_2))\right)\exp(-\kappa_{hA}z_2)P(\eta_2^2)dz_2;
\end{aligned}
\end{equation}
\begin{equation}\label{B2}
\begin{aligned}
&J_{12}=\frac{1}{2\rho_{max}k_{cA}}\sin(k_{cD}R_D)\exp(-\kappa_{cD}(d-R_D))\times\\
&\times\int_0^{R_A}\frac{1}{k_{cD}(d-z_1')}\exp(\kappa_{cD}z_1')dz_1'\times\\
&\times\left\{\left(\frac{\sin(k_{hD}R_D)}{k_{hD}R_D}-\cos(k_{hD}R_D)\right)\exp(-\kappa_{hD}(d-R_D))\right.\times\\
&\times\left[-2\int\limits_0^{z_1'}\frac{1}{k_{hD}(d-z_2)}\frac{1}{k_{hA}z_2}\right.\times\\
&\times\left.\left(\frac{k_{hA}z_2}{k_{hA}z_2}-\cos(k_{hA}z_2)\right)\exp(k_{hD}z_2)dz_2+\right.\\
&\left.+\int\limits_0^{R_A}\frac{1}{k_{hD}(d-z_2)}\frac{1}{k_{hA}z_2}\left(\frac{\sin(k_{hA}z_2)}{k_{hA}z_2}-\cos(k_{hA}z_2)\right)\exp(\kappa_{hD}z_2)dz_2\right]+\\
&+\left(\frac{\sin(k_{hD}R_D)}{k_{hD}R_D}-\cos(k_{hD}R_D)\right)\left(\frac{\sin(k_{hA}R_A)}{k_{hA}R_A}-\cos(k_{hA}R_A)\right)+\\
&+\exp(-\kappa_{hD}(d-R_D)+\kappa_{hA}R_A)\times\\
&\times\int\limits_{R_A}^{d-R_D}\frac{1}{k_{hD}(d-z_2)}\frac{1}{k_{hA}z_2}\exp((\kappa_{hD}-\kappa_{hA})z_2)dz_2+\\
&+\left(\frac{\sin(k_{hA}R_A)}{k_{hA}R_A}-\cos(k_{hA}R_A)\right)\exp(\kappa_{hA}R_A)\times\\
&\left.\times\int\limits_{d-R_D}^{d}\frac{1}{k_{hD}(d-z_2)}\frac{1}{k_{hA}z_2}\times\right.\\
&\times\left.\left(\frac{\sin(k_{hD}(d-z_2))}{k_{hD}(d-z_2)}-\cos(k_{hD}(d-z_2))\right)\exp(-\kappa_{hA}z_2)dz_2\right\};\\
\end{aligned}
\end{equation}
\begin{equation}\label{B3}
\begin{aligned}
&J_{13}=-\frac{1}{2\rho_{max}k_{hA}}\sin(k_{cD}R_D)\exp(-\kappa_{cD}(d-R_D))\int\limits_0^{R_A}\frac{1}{k_{cD}(d-z_1')}\frac{1}{k_{cA}z_1'}\sin(k_{cA}z_1')dz_1'\times\\
&\times\left\{\left(\frac{\sin(k_{hD}R_D)}{k_{hD}R_D}-\cos(k_{hD}R_D)\right)\right.\left.\exp(-\kappa_{hD}(d-R_D))\right.\times\\
&\times\left[-\int\limits_0^{z_1'}\frac{1}{k_{hD}(d-z_2)}\left(\frac{\sin(k_{hA}z_2)}{k_{hA}z_2}-\cos(k_{hA}z_2)\right)\exp(k_{hD}z_2)dz_2\right.+\\
&+\left(\frac{\sin(k_{hD}R_D)}{k_{hD}R_D}-\cos(k_{hD}R_D)\right)\left(\frac{\sin(k_{hA}R_A)}{k_{hA}R_A}-\cos(k_{hA}R_A)\right)\exp(-\kappa_{hD}(d-R_D)+\kappa_{hA}R_A)\times\\
&\times\left.\int\limits_{d-R_A}^{d}\frac{1}{k_{hD}(d-z_2)}\left(\frac{\sin(k_{hD}(d-z_2))}{k_{hD}(d-z_2)}-\cos(k_{hD}(d-z_2))\right)\exp(-\kappa_{hA}z_2)dz_2\right\}
\end{aligned}
\end{equation}
To calculate these integrals, the integrands containing sine and cosine functions are approximated with polynomials so that the error in calculating the integrals does not exceed 3\%.
Let us consider $J_{12}$:
\begin{equation}\label{B4}
\begin{aligned}
&J_{12}=\frac{1}{2\rho_{max}k_{cA}}\sin(k_{cD}R_D)\exp(-\kappa_{cD}(d-R_D))\int_0^{R_A}\frac{1}{k_{cD}(d-z_1')}\exp(\kappa_{cD}z_1')dz_1'\times\\
&\times\left\{\left(\frac{\sin(k_{hD}R_D)}{k_{hD}R_D}-\cos(k_{hD}R_D)\right)\exp(-\kappa_{hD}(d-R_D))\right.\times\\
&\times\left[-2\int\limits_0^{z_1'}\frac{1}{k_{hD}(d-z_2)}\frac{1}{k_{hA}z_2}\left(\frac{k_{hA}z_2}{k_{hA}z_2}-\cos(k_{hA}z_2)\right)\exp(k_{hD}z_2)dz_2+\right.\\
&\left.+\int\limits_0^{R_A}\frac{1}{k_{hD}(d-z_2)}\frac{1}{k_{hA}z_2}\left(\frac{\sin(k_{hA}z_2)}{k_{hA}z_2}-\cos(k_{hA}z_2)\right)\exp(\kappa_{hD}z_2)dz_2\right]+\\
&+\left(\frac{\sin(k_{hD}R_D)}{k_{hD}R_D}-\cos(k_{hD}R_D)\right)\left(\frac{\sin(k_{hA}R_A)}{k_{hA}R_A}-\cos(k_{hA}R_A)\right)\exp(-\kappa_{hD}(d-R_D)+\kappa_{hA}R_A)\times\\
&\times\int\limits_{R_A}^{d-R_D}\frac{1}{k_{hD}(d-z_2)}\frac{1}{k_{hA}z_2}\exp((\kappa_{hD}-\kappa_{hA})z_2)dz_2+\left(\frac{\sin(k_{hA}R_A)}{k_{hA}R_A}-\cos(k_{hA}R_A)\right)\exp(\kappa_{hA}R_A)\times\\
&\left.\times\int\limits_{d-R_D}^{d}\frac{1}{k_{hD}(d-z_2)}\frac{1}{k_{hA}z_2}\left(\frac{\sin(k_{hD}(d-z_2))}{k_{hD}(d-z_2)}-\cos(k_{hD}(d-z_2))\right)\exp(-\kappa_{hA}z_2)dz_2\right\}.\\
\end{aligned}                                                                             
\end{equation}
In the integrals enclosed in square brackets, we change variables: $k_{kA}z_2=y$.
Then, the first of the integrals in square brackets is transformed to
$\frac{1}{k_{hD}d}\frac{1}{k_{hA}}\int\limits_0^{k_{hA}z_1'}\frac{1}{1-y/(k_{hA}d)}\frac{1}{y}\left(\frac{\sin y}{y}-\cos y\right)\exp(a y)dy $
where the designation $a=k_{hD}/k_{hA}$ is introduced.
The replacement in this integral of the function $\frac{1}{1-y/(k_{hA}d)}\frac{1}{y}\left(\frac{\sin y}{y}-\cos y\right)$ with the polynomial
$\frac{3}{8}y\left(1+\frac{y}{k_{hD}d}-\left(\frac{y}{k_{hD}d}\right)^2\right)\left(1-\frac{y}{k_{hD}R_A}\right)$
makes it possible to easily calculate the integral to obtain
\begin{equation}\label{B5}
\begin{aligned}
&\frac{1}{k_{hA}}\frac{1}{k_{hD}d}\exp(k_{hD}z_1')\frac{3}{8}\frac{1}{a^2}\left\{\left[\kappa_{hD}z_1'-1+\exp(-\kappa_{hD}z_1')\right]-\right.\\
&-\frac{\alpha_A}{a}\left[(\kappa_{hD}z_1')^2-2(\kappa_{hD}z_1')+2-2\exp(-\kappa_{hD}z_1')\right]-\\
&-\frac{\beta_A}{a^2}\left[(\kappa_{hD}z_1')^3-3(\kappa_{hD}z_1')^2+6(\kappa_{hD}z_1')-6+6\exp(-\kappa_{hD}z_1')\right]+\\
&\left.+\frac{\gamma_A}{a^3}\left[(\kappa_{hD}z_1')^4-4(\kappa_{hD}z_1')^3+12(\kappa_{hD}z_1')^2-24(\kappa_{hD}z_1')+24-24\exp(-\kappa_{hD}z_1')\right]\right\},
\end{aligned}
\end{equation}
where $\alpha_A=(1/(k_{hD}R_A))-(1/(k_{hD}d))$, $\beta_A=(1/(k_{hD}d))((1/(k_{hD}R_A))+(1/(k_{hD}d)))$,
$\gamma_A=(1/(k_{hD}d)^2)(1/(k_{hD}R_A))$.
The second integral in square brackets is calculated in a similar way:
\begin{equation}\label{B6}
\begin{aligned}
&\frac{1}{k_{hA}}\frac{1}{k_{hD}d}\exp(\kappa_{hD}R_A)\frac{3}{8}\frac{1}{a^2}\left\{(\kappa_{hD}R_A)-1-\right.\\
&-\frac{\alpha_A}{a}\left[(\kappa_{hD}R_A)^2-2(\kappa_{hD}R_A)+2\right]-\frac{\beta_A}{a^2}\left[(\kappa_{hD}R_A)^3-3(\kappa_{hD}R_A)^2+6(\kappa_{hD}R_A)-6\right]+\\
&+\left.\frac{\gamma_A}{a^3}\left[(\kappa_{hD}R_A)^4-4(\kappa_{hD}R_A)^3+12(\kappa_{hD}R_A)^2-24(\kappa_{hD}R_A)+24\right]\right\}.
\end{aligned}
\end{equation}
After the change of variables $k_{cA}z_1'=y$, the integral
$\int\limits_0^{R_A}\frac{1}{k_{cD}(d-z_1')}\sin(k_{cA}z_1')\exp(\kappa_{cD}z_1')dz_1'$ is brought to the form
$\frac{1}{k_{cD}d}\frac{1}{k_{cA}}\int\limits_0^{k_{cA}R_A}\frac{1}{1-y/(k_{cA}d)}\sin(y)\exp(by)$
where $b=\kappa_{cD}/\kappa_{cA}$. In this case, the integrand $\frac{\sin(y)}{1-y/(k_{cD}d)}$  is modeled by the expression
$y(1+y/(k_{cD}d)+y^2/(k_{cD}d))(1-y/\pi)$
As a result, we have
\begin{equation}\label{B7}
\begin{aligned}
&\int\limits_0^{R_A}\frac{1}{k_{cD}(d-z_1')}\sin(k_{cA}z_1')\exp(\kappa_{cD}z_1')dz_1'=\frac{1}{k_{cD}d}\frac{1}{k_{cA}}\exp(-\kappa_{cD}R_A)\times\\
&\times\frac{1}{b^2}\left\{\left[(\kappa_{cD}R_A-1)+\exp(-\kappa_{cD}R_A)\right]-\left(\frac{1}{\pi}-\frac{1}{k_{cA}d}\right)\right.\times\\
&\times\left.\frac{1}{b}\left[(\kappa_{cD}R_A)^2-2(\kappa_{cD}R_A)+2-2\exp(-\kappa_{cD}R_A)\right]\right.+\\
&+\left(1-\frac{1}{\pi}\right)\frac{1}{b^2}\left[(\kappa_{cD}R_A)^3-3(\kappa_{cD}R_A)^2+6(\kappa_{cD}R_A)-6+6\exp(-\kappa_{cD}R_A)\right]-\\
&\left.-\frac{1}{\pi}\frac{1}{k_{cA}d}\frac{1}{b^3}\left[(\kappa_{cD}R_A)^4-4(\kappa_{cD}R_A)^3+12(\kappa_{cD}R_A)^2-24(\kappa_{cD}R_A)+24-24\exp(-\kappa_{cD}R_A)\right]\right\}
\end{aligned}
\end{equation}
Substitution of expression (\ref{B3}) into $\int\limits_0^{R_A}f(z_1')dz_1'$ (\ref{B2}),
replacement of variables $k_{cA}z_1'=y$, and use of the auxiliary function $\left(\frac{k_{hA}R_A}{10}y^2\left(1-\frac{y^2}{k_{hA}R_A}\right)\right)$  makes it possible to calculate this integral and obtain
\begin{equation}\label{B8}
\begin{aligned}
&\frac{1}{k_{cA}}\frac{1}{k_{cD}d}\frac{1}{k_{hA}}\frac{1}{k_{hD}d}\exp((\kappa_{cD}+\kappa_{hD})R_A)\times\\
&\times\frac{k_{hA}R_{A}}{10}\frac{1}{p^3}\left\{\left[(pk_{cA}R_{A})^2-2(pk_{cA}R_{A})+2\right]\right.-\\
&\left.-\frac{1}{k_{hA}R_A}\frac{1}{p^2}\left[(pk_{cA}R_{A})^4-4(pk_{cA}R_{A})^3+12(pk_{cA}R_{A})^2-24(pk_{cA}R_{A})+24\right]\right\},
\end{aligned}
\end{equation}
where  $p=(\kappa_{cD}+\kappa_{hD})/k_{cA}$
The integral over the region between the QDs, $R_A \Longleftrightarrow d-R_D$ , in (\ref{B1}) is easily calculated if we take into account that
\begin{equation}\label{B9}
\begin{aligned}
&\kappa_{hA}\cong \kappa_{hD}: \, \int\limits_{R_A}^{d-R_D}\frac{1}{k_{hD}(d-z_2)}\frac{1}{k_{hA}z_2}\exp((\kappa_{hD}-\kappa_{hA})z_2)dz_2\cong\\
&\cong \frac{1}{k_{hD}d}\frac{1}{k_{hA}}\int\limits_{R_A}^{d-R_D}\left(\frac{1}{z_2}+\frac{1}{d-z_2}\right)=\frac{1}{k_{hD}d}\frac{1}{k_{hA}}\ln \frac{(d-R_D)(d-R_A)}{R_A R_D}
\end{aligned}
\end{equation}
Calculation of the integral over the donor region:
\begin{equation}\label{B10}
\int\limits_{d-R_D}^d\frac{1}{k_{hD}(d-z_2)}\frac{1}{k_{hA}z_2}\left(\frac{\sin (k_{hD}(d-z_2))}{k_{hD}(d-z_2)}-\cos(k_{hD}(d-z_2))\right)\exp(-\kappa_{hA}z_2)dz_2.
\end{equation}
By the replacement of variables:  $k_{hD}(d-z_2)=y$ the integral is
brought to the form $$\frac{1}{k_{hD}d}\frac{1}{k_{hA}}\exp(-\kappa_{hA}d)\int\limits_0^{k_{hD}R_D}\frac{1}{1-y/(k_{hD}d)}\frac{1}{y}\left(\frac{\sin(y)}{y}-\cos(y)\right)\exp(cy)dy$$ ($c=\kappa_{hD}/\kappa_{hA}$),
considered above. As a result, we come to expression (\ref{B4}), in which it is necessary to change in the literal designations the indexes A for D.
Let us write the final expression for integral (\ref{B2}):
\begin{equation}\label{B11}                   
\begin{aligned}
&J_{12}=\frac{1}{2\rho_{max}k_{cA}}\sin(k_{cD}R_D)\frac{1}{k_{cD}d}\frac{1}{k_{cA}}\frac{1}{k_{hD}d}\frac{1}{k_{hA}}\exp(-(\kappa_{cD}+\kappa_{hD})(d-R_D-R_A))\times\\
&\times\left\{\left(\frac{\sin(k_{hD}R_D)}{(k_{hD}R_D)}-\cos(k_{hD}R_D)\right)(B_A'+B_A)\right.+\\
&+\left(\frac{\sin(k_{hD}R_D)}{(k_{hD}R_D)}-\cos(k_{hD}R_D)\right)\left(\frac{\sin(k_{hA}R_A)}{(k_{hA}R_A)}-\cos(k_{hA}R_A)\right)B_{A-D}+\\
&\left.+\left(\frac{\sin(k_{hA}R_A)}{(k_{hA}R_A)}-\cos(k_{hA}R_A)\right)B_D\right\}
\end{aligned}
\end{equation}
where 
\begin{equation}\label{B12}
B_A'=-2\left(\frac{2}{k_{hD}R_A}\frac{1}{a^2}\right)\left(\frac{3}{4}\frac{1}{\gamma^2}\right)\left\{3(\gamma k_{cA}R_A-1)-\frac{1}{\gamma}\left[(\gamma k_{cA}R_A)^2-2(\gamma k_{cA}R_A)+2\right]\right\}.
\end{equation}
Here $a=\kappa_{hD}/k_{hA}$, $\gamma=(\kappa_{cD}+\kappa_{hD})/k_{cA}$;
\begin{equation}\label{B13}
\begin{aligned}
&B_A=\left(\frac{2}{\kappa_{hD}R_A}\frac{1}{a^2}\right)\left\{(\kappa_{hD}R_A)^2-2(\kappa_{hD}R_A)+2\right\}\times\\
&\times\left(\frac{1}{b^2}\right)\left\{\left[(\kappa_{cD}R_A-1)+\exp(-\kappa_{cD}R_A)\right]-\left(\frac{1}{\pi}-\frac{1}{k_{cD}d}\right)\frac{1}{b}\left[(\kappa_{cD}R_A)^2-2(\kappa_{cD}R_A)+2-2(-\kappa_{cD}R_A)\right]\right.+\\
&+\left(1-\frac{1}{\pi}\right)\frac{1}{k_{cD}d}\frac{1}{b^2}\left[(\kappa_{cD}R_A)^3-3(\kappa_{cD}R_A)^2+6(\kappa_{cD}R_A)-6+6\exp(-\kappa_{cD}R_A)\right]-\\
&\left.-\frac{1}{\pi}\frac{1}{k_{cD}d}\frac{1}{b^3}\left[(\kappa_{cD}R_A)^4-4(\kappa_{cD}R_A)^3+12(\kappa_{cD}R_A)^2-24(\kappa_{cD}R_A)+24-24\exp(-\kappa_{cD}R_A)\right]\right\},
\end{aligned}
\end{equation}
where  $b=\kappa_{cD}/k_{cA}$.
\begin{equation}\label{B14}
\begin{aligned}
&B_{A-D}=\left(\frac{1}{b^2}\right)\left\{\left[(\kappa_{cD}R_A)-1+\exp(-\kappa_{cD}R_A)\right]-\left(\frac{1}{\pi}-\frac{1}{k_{cA}d}\right)\right.\times\\
&\times\left.\frac{1}{b}\left[(\kappa_{cD}R_A)^2-2(\kappa_{cD}R_A)+2-2\exp(-\kappa_{cD}R_A)\right]\right.+\\
&+\left(1-\frac{1}{\pi}\right)\frac{1}{b^2}\left[(\kappa_{cD}R_A)^3-3(\kappa_{cD}R_A)^2+6(\kappa_{cD}R_A)-6+6\exp(-\kappa_{cD}R_A)\right]-\\
&\left.-\frac{1}{\pi}\frac{1}{k_{cA}d}\frac{1}{b^3}\left[(\kappa_{cD}R_A)^4-4(\kappa_{cD}R_A)^3+12(\kappa_{cD}R_A)^2-24(\kappa_{cD}R_A)+24-24\exp(-\kappa_{cD}R_A)\right]\right\}\times\\
&\times\ln\frac{(d-R_A)(d-R_D)}{R_A R_D}.
\end{aligned}
\end{equation}
In this expression, account is taken of the fact that  $\kappa_{hA}\cong\kappa_{hD}$.
\begin{equation}\label{B15}
\begin{aligned}
&B_{A-D}=\left(\frac{1}{b^2}\right)\times\left\{\left[(\kappa_{cD}R_A)-1+\exp(-\kappa_{cD}R_A)\right]-\left(\frac{1}{\pi}-\frac{1}{k_{cA}d}\right)\right.\times\\
&\times\left.\frac{1}{b}\left[(\kappa_{cD}R_A)^2-2(\kappa_{cD}R_A)+2-2\exp(-\kappa_{cD}R_A)\right]\right.+\\
&+\left(1-\frac{1}{\pi}\right)\frac{1}{b^2}\left[(\kappa_{cD}R_A)^3-3(\kappa_{cD}R_A)^2+6(\kappa_{cD}R_A)-6+6\exp(-\kappa_{cD}R_A)\right]-\\
&-\frac{1}{\pi}\frac{1}{k_{cA}d}\frac{1}{b^3}\left[(\kappa_{cD}R_A)^4-4(\kappa_{cD}R_A)^3+12(\kappa_{cD}R_A)^2-24(\kappa_{cD}R_A)+\right.\\
&+\left.24-24\exp(-\kappa_{cD}R_A)\right]\times\left(\frac{2}{k_{hD}R_D}\frac{1}{c^2}\right)\left\{(\kappa_{cD}R_A)-1-\frac{1}{(\kappa_{cD}R_A)}\right.\times\\
&\times\left.\left[(\kappa_{cD}R_A)^2-2(\kappa_{cD}R_A)+2\right]\right\},
\end{aligned}
\end{equation}
here  $c=\kappa_{hD}/k_{hA}$.
The integrals appearing in $J_2$, $J_3$ are calculated in a similar manner. The substitution of the system parameter numerical values in the obtained expressions results in (46).
Two recent papers \cite{add04,add05}
support our result that in the systems including QDs the exchange interaction can play a significant role in the energy transfer
at small donor-acceptor separation. It is experimentally shown in
\cite{add04}
that in the CdSe QD-Squaraine light harvesting assemblies, the exchange (Dexter) process is essential to the energy
transfer and dominates the dipole-based Forster one at the smaller QDs. In
\cite{add05}
the measured energy transfer rates in the close-packed blends of  CdSe/CdZnS core/shell QDs are found to be more than an
order of magnitude larger than the rate predicted by Forster theory,  which cannot be satisfactory explained by several
possible factors considered in the paper. We believe that exchange contribution to the energy transfer should be also
taken into account for the consideration the discrepancy.

\begin{mcitethebibliography}{68}
\providecommand*\natexlab[1]{#1}
\providecommand*\mciteSetBstSublistMode[1]{}
\providecommand*\mciteSetBstMaxWidthForm[2]{}
\providecommand*\mciteBstWouldAddEndPuncttrue
  {\def\EndOfBibitem{\unskip.}}
\providecommand*\mciteBstWouldAddEndPunctfalse
  {\let\EndOfBibitem\relax}
\providecommand*\mciteSetBstMidEndSepPunct[3]{}
\providecommand*\mciteSetBstSublistLabelBeginEnd[3]{}
\providecommand*\EndOfBibitem{}
\mciteSetBstSublistMode{f}
\mciteSetBstMaxWidthForm{subitem}{(\alph{mcitesubitemcount})}
\mciteSetBstSublistLabelBeginEnd
  {\mcitemaxwidthsubitemform\space}
  {\relax}
  {\relax}

\bibitem[Agranovich and Galanin(1983)Agranovich, and Galanin]{ref01}
Agranovich,~V.~M.; Galanin,~M.~D. \emph{Electronic Excitation Energy Transfer
  in Condensed Matter}; North-Holland: Amsterdam, The Netherlands, 1983\relax
\mciteBstWouldAddEndPuncttrue
\mciteSetBstMidEndSepPunct{\mcitedefaultmidpunct}
{\mcitedefaultendpunct}{\mcitedefaultseppunct}\relax
\EndOfBibitem
\bibitem[Andrews(1989)]{ref02}
Andrews,~D.~L. \emph{Chem. Phys} \textbf{1989}, \emph{135}, 195--201\relax
\mciteBstWouldAddEndPuncttrue
\mciteSetBstMidEndSepPunct{\mcitedefaultmidpunct}
{\mcitedefaultendpunct}{\mcitedefaultseppunct}\relax
\EndOfBibitem
\bibitem[Scholes and Andrews(2005)Scholes, and Andrews]{ref03}
Scholes,~G.~D.; Andrews,~D.~L. Resonance energy transfer and quantum dots.
  \emph{Phys. Rev. B} \textbf{2005}, \emph{72}, 125331\relax
\mciteBstWouldAddEndPuncttrue
\mciteSetBstMidEndSepPunct{\mcitedefaultmidpunct}
{\mcitedefaultendpunct}{\mcitedefaultseppunct}\relax
\EndOfBibitem
\bibitem[Forster(1948)]{ref04}
Forster,~T. Zwischenmolekulare Energiewanderung und Fluoreszenz. \emph{Ann.
  Phys.} \textbf{1948}, \emph{437}, 55\relax
\mciteBstWouldAddEndPuncttrue
\mciteSetBstMidEndSepPunct{\mcitedefaultmidpunct}
{\mcitedefaultendpunct}{\mcitedefaultseppunct}\relax
\EndOfBibitem
\bibitem[Dexter(1953)]{ref05}
Dexter,~D.~L. A Theory of Sensitized Luminescence in Solids. \emph{J. Chem.
  Phys.} \textbf{1953}, \emph{21}, 836\relax
\mciteBstWouldAddEndPuncttrue
\mciteSetBstMidEndSepPunct{\mcitedefaultmidpunct}
{\mcitedefaultendpunct}{\mcitedefaultseppunct}\relax
\EndOfBibitem
\bibitem[Mulliken(1950)]{ref06}
Mulliken,~R.~S. Structures of Complexes Formed by Halogen Molecules with
  Aromatic and with Oxygenated Solvents. \emph{J. Am. Chem. Soc.}
  \textbf{1950}, \emph{72}, 600\relax
\mciteBstWouldAddEndPuncttrue
\mciteSetBstMidEndSepPunct{\mcitedefaultmidpunct}
{\mcitedefaultendpunct}{\mcitedefaultseppunct}\relax
\EndOfBibitem
\bibitem[Cario and Franck(1923)Cario, and Franck]{ref07}
Cario,~G.; Franck,~J. Uber sensibilisierte fluoreszenz von gasen. \emph{Z.
  Physik} \textbf{1923}, \emph{17}, 202\relax
\mciteBstWouldAddEndPuncttrue
\mciteSetBstMidEndSepPunct{\mcitedefaultmidpunct}
{\mcitedefaultendpunct}{\mcitedefaultseppunct}\relax
\EndOfBibitem
\bibitem[Terenin and Karyakin(1951)Terenin, and Karyakin]{ref08}
Terenin,~A.~N.; Karyakin,~A.~V. \emph{Izv. Akad. Nauk SSSR, Ser. Fiz}
  \textbf{1951}, \emph{15}, 550\relax
\mciteBstWouldAddEndPuncttrue
\mciteSetBstMidEndSepPunct{\mcitedefaultmidpunct}
{\mcitedefaultendpunct}{\mcitedefaultseppunct}\relax
\EndOfBibitem
\bibitem[Perrin and Chocroun(1929)Perrin, and Chocroun]{ref09}
Perrin,~J.; Chocroun,~C.~R. \emph{habd. Acad. Sci. Seances} \textbf{1929},
  \emph{189}, 1213\relax
\mciteBstWouldAddEndPuncttrue
\mciteSetBstMidEndSepPunct{\mcitedefaultmidpunct}
{\mcitedefaultendpunct}{\mcitedefaultseppunct}\relax
\EndOfBibitem
\bibitem[Forster(1949)]{ref10}
Forster,~T. Versuche zum zwischenmolekularen Ubergang von
  Electroneneanregungsenergie. \emph{Z. Electrochem} \textbf{1949}, \emph{53},
  93\relax
\mciteBstWouldAddEndPuncttrue
\mciteSetBstMidEndSepPunct{\mcitedefaultmidpunct}
{\mcitedefaultendpunct}{\mcitedefaultseppunct}\relax
\EndOfBibitem
\bibitem[Galanin and Levshin(1951)Galanin, and Levshin]{ref11}
Galanin,~M.~D.; Levshin,~V.~L. \emph{Zh. Eksp. Teor. Fiz.} \textbf{1951},
  \emph{21}, 121\relax
\mciteBstWouldAddEndPuncttrue
\mciteSetBstMidEndSepPunct{\mcitedefaultmidpunct}
{\mcitedefaultendpunct}{\mcitedefaultseppunct}\relax
\EndOfBibitem
\bibitem[Terenin and Ermolaev(1951)Terenin, and Ermolaev]{ref12}
Terenin,~A.~N.; Ermolaev,~V.~L. \emph{Dokl. Akad. Nauk. SSSR} \textbf{1951},
  \emph{85}, 547\relax
\mciteBstWouldAddEndPuncttrue
\mciteSetBstMidEndSepPunct{\mcitedefaultmidpunct}
{\mcitedefaultendpunct}{\mcitedefaultseppunct}\relax
\EndOfBibitem
\bibitem[Emerson and Arnold(1932)Emerson, and Arnold]{ref13}
Emerson,~R.; Arnold,~W. A separation of the reactions in photosynthesis by
  means of intermittent light. \emph{J. Gen. Physiol.} \textbf{1932},
  \emph{16}, 191\relax
\mciteBstWouldAddEndPuncttrue
\mciteSetBstMidEndSepPunct{\mcitedefaultmidpunct}
{\mcitedefaultendpunct}{\mcitedefaultseppunct}\relax
\EndOfBibitem
\bibitem[Scholes(2003)]{ref14}
Scholes,~G.~D. Long-range resonance energy transfer in molecular systems.
  \emph{Annu. Rev. Phys. Chem.} \textbf{2003}, \emph{54}, 57\relax
\mciteBstWouldAddEndPuncttrue
\mciteSetBstMidEndSepPunct{\mcitedefaultmidpunct}
{\mcitedefaultendpunct}{\mcitedefaultseppunct}\relax
\EndOfBibitem
\bibitem[Ha \latin{et~al.}(1996)Ha, Enderle, Ogletree, Chemla, Selvin, and
  Weiss]{ref15}
Ha,~T.; Enderle,~T.; Ogletree,~D.~F.; Chemla,~D.~S.; Selvin,~P.~R.; Weiss,~S.
  Probing the interaction between two single molecules: fluorescence resonance
  energy transfer between a single donor and a single acceptor. Proc. Natl.
  Acad. Sci. USA, Biophysics. 1996; p 6264\relax
\mciteBstWouldAddEndPuncttrue
\mciteSetBstMidEndSepPunct{\mcitedefaultmidpunct}
{\mcitedefaultendpunct}{\mcitedefaultseppunct}\relax
\EndOfBibitem
\bibitem[Kao \latin{et~al.}(2008)Kao, Yang, Lin, Lim, Fann, and Chen]{ref16}
Kao,~M. W.-P.; Yang,~L.-L.; Lin,~J. C.-K.; Lim,~T.-S.; Fann,~W.; Chen,~R. P.-Y.
  Strategy for Efficient Site-Specific FRET-Dye Labeling of Ubiquitin.
  \emph{Bioconjugate Chem.} \textbf{2008}, \emph{19}, 1124\relax
\mciteBstWouldAddEndPuncttrue
\mciteSetBstMidEndSepPunct{\mcitedefaultmidpunct}
{\mcitedefaultendpunct}{\mcitedefaultseppunct}\relax
\EndOfBibitem
\bibitem[Kagan \latin{et~al.}(1996)Kagan, Murray, Nirmal, and Bawendi]{ref17}
Kagan,~C.~R.; Murray,~C.~B.; Nirmal,~M.; Bawendi,~M.~J. Electronic Energy
  Transfer in CdSe Quantum Dot Solids. \emph{Phys. Rev. Lett.} \textbf{1996},
  \emph{76}, 1517\relax
\mciteBstWouldAddEndPuncttrue
\mciteSetBstMidEndSepPunct{\mcitedefaultmidpunct}
{\mcitedefaultendpunct}{\mcitedefaultseppunct}\relax
\EndOfBibitem
\bibitem[Clapp \latin{et~al.}(2006)Clapp, Medintz, and Mattousi]{ref18}
Clapp,~A.~R.; Medintz,~I.~L.; Mattousi,~H. Forster resonance energy transfer
  investigations using quantum-dot fluorophores. \emph{Chem. Phys. Chem.}
  \textbf{2006}, \emph{7}, 47\relax
\mciteBstWouldAddEndPuncttrue
\mciteSetBstMidEndSepPunct{\mcitedefaultmidpunct}
{\mcitedefaultendpunct}{\mcitedefaultseppunct}\relax
\EndOfBibitem
\bibitem[Delerue and Allan(2007)Delerue, and Allan]{ref19}
Delerue,~C.; Allan,~G. Energy transfer between semiconductor nanocrystals:
  Validity of Forsters theory. \emph{Phys. Rev. B} \textbf{2007}, \emph{75},
  195311\relax
\mciteBstWouldAddEndPuncttrue
\mciteSetBstMidEndSepPunct{\mcitedefaultmidpunct}
{\mcitedefaultendpunct}{\mcitedefaultseppunct}\relax
\EndOfBibitem
\bibitem[Curutchet \latin{et~al.}(2008)Curutchet, Franceschetti, and
  Zunger]{ref20}
Curutchet,~C.; Franceschetti,~A.; Zunger,~A. Examining Forster Energy Transfer
  for Semiconductor Nanocrystalline Quantum Dot Donors and Acceptors. \emph{J.
  Phys. Chem. C.} \textbf{2008}, \emph{112}, 13336\relax
\mciteBstWouldAddEndPuncttrue
\mciteSetBstMidEndSepPunct{\mcitedefaultmidpunct}
{\mcitedefaultendpunct}{\mcitedefaultseppunct}\relax
\EndOfBibitem
\bibitem[Baer and Rabani(2008)Baer, and Rabani]{ref21}
Baer,~R.; Rabani,~E. Theory of resonance energy transfer involving
  nanocrystals: The role of high multipoles. \emph{J. Chem. Phys.}
  \textbf{2008}, \emph{128}, 184710\relax
\mciteBstWouldAddEndPuncttrue
\mciteSetBstMidEndSepPunct{\mcitedefaultmidpunct}
{\mcitedefaultendpunct}{\mcitedefaultseppunct}\relax
\EndOfBibitem
\bibitem[Kruchinin \latin{et~al.}(2008)Kruchinin, Fedorov, Baranov, Perova, and
  Berwick]{ref22}
Kruchinin,~S.~Y.; Fedorov,~A.~V.; Baranov,~A.~N.; Perova,~S.; Berwick,~K.
  Resonant energy transfer in quantum dots: Frequency-domain luminescent
  spectroscopy. \emph{Phys. Rev. B} \textbf{2008}, \emph{78}, 125311\relax
\mciteBstWouldAddEndPuncttrue
\mciteSetBstMidEndSepPunct{\mcitedefaultmidpunct}
{\mcitedefaultendpunct}{\mcitedefaultseppunct}\relax
\EndOfBibitem
\bibitem[Medintz and Mattoussi(2009)Medintz, and Mattoussi]{ref23}
Medintz,~I.~L.; Mattoussi,~H. \emph{Phys. Chem. Chem. Phys.} \textbf{2009},
  \emph{11}, 17--45\relax
\mciteBstWouldAddEndPuncttrue
\mciteSetBstMidEndSepPunct{\mcitedefaultmidpunct}
{\mcitedefaultendpunct}{\mcitedefaultseppunct}\relax
\EndOfBibitem
\bibitem[Chaniotakis and Frasco(2010)Chaniotakis, and Frasco]{ref24}
Chaniotakis,~N.; Frasco,~M.~F. \emph{Anal. Bioanal. Chem.} \textbf{2010},
  \emph{396}, 229--240\relax
\mciteBstWouldAddEndPuncttrue
\mciteSetBstMidEndSepPunct{\mcitedefaultmidpunct}
{\mcitedefaultendpunct}{\mcitedefaultseppunct}\relax
\EndOfBibitem
\bibitem[Chou and Dennis(2015)Chou, and Dennis]{ref25}
Chou,~K.~F.; Dennis,~A.~M. \emph{Sensors} \textbf{2015}, \emph{15},
  13288--13325\relax
\mciteBstWouldAddEndPuncttrue
\mciteSetBstMidEndSepPunct{\mcitedefaultmidpunct}
{\mcitedefaultendpunct}{\mcitedefaultseppunct}\relax
\EndOfBibitem
\bibitem[Hildebrandt \latin{et~al.}(2017)Hildebrandt, Spillman, Algar, Pons,
  Stewart, Oh, Susumu, Diaz, Delehanty, and Medintz]{ref26}
Hildebrandt,~N.; Spillman,~C.~M.; Algar,~W.~R.; Pons,~T.; Stewart,~M.~H.;
  Oh,~E.; Susumu,~K.; Diaz,~S.~A.; Delehanty,~J.~B.; Medintz,~I.~L. \emph{Chem.
  Rev.} \textbf{2017}, \emph{117}, 536--711\relax
\mciteBstWouldAddEndPuncttrue
\mciteSetBstMidEndSepPunct{\mcitedefaultmidpunct}
{\mcitedefaultendpunct}{\mcitedefaultseppunct}\relax
\EndOfBibitem
\bibitem[Lovett \latin{et~al.}(2003)Lovett, Reina, Nazir, and Breggs]{ref27}
Lovett,~B.~W.; Reina,~J.~H.; Nazir,~A.; Breggs,~A.~D. Optical schemes for
  quantum computation in quantum dot molecules. \emph{Phys. Rev. B}
  \textbf{2003}, \emph{68}, 205319\relax
\mciteBstWouldAddEndPuncttrue
\mciteSetBstMidEndSepPunct{\mcitedefaultmidpunct}
{\mcitedefaultendpunct}{\mcitedefaultseppunct}\relax
\EndOfBibitem
\bibitem[Noda(2006)]{ref28}
Noda,~S. Seeking the Ultimate Nanolaser. \emph{Science} \textbf{2006},
  \emph{314}, 260\relax
\mciteBstWouldAddEndPuncttrue
\mciteSetBstMidEndSepPunct{\mcitedefaultmidpunct}
{\mcitedefaultendpunct}{\mcitedefaultseppunct}\relax
\EndOfBibitem
\bibitem[Heitz \latin{et~al.}(1999)Heitz, Mukhamedov, Zeng, Chen, Madhukar, and
  Bimberg]{ref29}
Heitz,~R.; Mukhamedov,~I.; Zeng,~J.; Chen,~P.; Madhukar,~A.; Bimberg,~D.
  Excitation transfer in novel self-organized quantum dot structures.
  \emph{Superlattices Microstruct.} \textbf{1999}, \emph{25}, 97\relax
\mciteBstWouldAddEndPuncttrue
\mciteSetBstMidEndSepPunct{\mcitedefaultmidpunct}
{\mcitedefaultendpunct}{\mcitedefaultseppunct}\relax
\EndOfBibitem
\bibitem[Law \latin{et~al.}(2008)Law, Luther, Song, Hughes, Perkins, and
  Nozik]{ref30}
Law,~M.; Luther,~J.~M.; Song,~O.; Hughes,~B.~R.; Perkins,~C.~L.; Nozik,~A.~J.
  Structural, Optical, and Electrical Properties of PbSe Nanocrystal Solids
  Treated Thermally or with Simple Amines. \emph{J. Am. Chem. Soc.}
  \textbf{2008}, \emph{130}, 5974\relax
\mciteBstWouldAddEndPuncttrue
\mciteSetBstMidEndSepPunct{\mcitedefaultmidpunct}
{\mcitedefaultendpunct}{\mcitedefaultseppunct}\relax
\EndOfBibitem
\bibitem[Choi \latin{et~al.}(2016)Choi, Wang, Oh, Paik, Sung, Sung, Ye, Zhao,
  Diroll, Murray, and Kagan]{add01}
Choi,~J.~H.; Wang,~H.; Oh,~S.~J.; Paik,~T.; Sung,~P.; Sung,~J.; Ye,~X.;
  Zhao,~T.; Diroll,~B.~T.; Murray,~C.~B. \latin{et~al.}  \emph{Science}
  \textbf{2016}, \emph{352}, 205--208\relax
\mciteBstWouldAddEndPuncttrue
\mciteSetBstMidEndSepPunct{\mcitedefaultmidpunct}
{\mcitedefaultendpunct}{\mcitedefaultseppunct}\relax
\EndOfBibitem
\bibitem[Bodunov and Shekhtman(1970)Bodunov, and Shekhtman]{ref31}
Bodunov,~E.~N.; Shekhtman,~V.~L. \emph{Sov. Phys. Solid State} \textbf{1970},
  \emph{12}, 2809\relax
\mciteBstWouldAddEndPuncttrue
\mciteSetBstMidEndSepPunct{\mcitedefaultmidpunct}
{\mcitedefaultendpunct}{\mcitedefaultseppunct}\relax
\EndOfBibitem
\bibitem[Agranovich \latin{et~al.}(1997)Agranovich, Rossa, and Bassani]{ref32}
Agranovich,~V.~M.; Rossa,~G. C.~L.; Bassani,~F. Efficient electronic energy
  transfer from a semiconductor quantum well to an organic material.
  \emph{ZhETF} \textbf{1997}, \emph{66}, 714\relax
\mciteBstWouldAddEndPuncttrue
\mciteSetBstMidEndSepPunct{\mcitedefaultmidpunct}
{\mcitedefaultendpunct}{\mcitedefaultseppunct}\relax
\EndOfBibitem
\bibitem[Agranovich and Basko(1999)Agranovich, and Basko]{ref33}
Agranovich,~V.~M.; Basko,~D.~M. \emph{JETP Lett.} \textbf{1999}, \emph{69},
  250\relax
\mciteBstWouldAddEndPuncttrue
\mciteSetBstMidEndSepPunct{\mcitedefaultmidpunct}
{\mcitedefaultendpunct}{\mcitedefaultseppunct}\relax
\EndOfBibitem
\bibitem[Basko \latin{et~al.}(1999)Basko, Rossa, Bassani, and
  Agranovich]{ref34}
Basko,~D.; Rossa,~G. C.~L.; Bassani,~F.; Agranovich,~V. Forster energy transfer
  from a semiconductor quantum well to an organic material overlayer.
  \emph{Eur. Phys. J. B.} \textbf{1999}, \emph{8}, 353\relax
\mciteBstWouldAddEndPuncttrue
\mciteSetBstMidEndSepPunct{\mcitedefaultmidpunct}
{\mcitedefaultendpunct}{\mcitedefaultseppunct}\relax
\EndOfBibitem
\bibitem[Basko \latin{et~al.}(2000)Basko, Agranovich, Bassani, and
  Rossa]{ref35}
Basko,~D.~M.; Agranovich,~V.~M.; Bassani,~F.; Rossa,~G. C.~L. Energy transfer
  from a semiconductor quantum dot to an organic matrix. \emph{Eur. Phys. J. B}
  \textbf{2000}, \emph{13}, 653\relax
\mciteBstWouldAddEndPuncttrue
\mciteSetBstMidEndSepPunct{\mcitedefaultmidpunct}
{\mcitedefaultendpunct}{\mcitedefaultseppunct}\relax
\EndOfBibitem
\bibitem[Agranovich \latin{et~al.}(2011)Agranovich, Gardstein, and
  Litinskaya]{ref36}
Agranovich,~V.~M.; Gardstein,~Y.~N.; Litinskaya,~M. Hybrid Resonant
  OrganicInorganic Nanostructures for Optoelectronic Applications. \emph{J.
  Chem. Rev.} \textbf{2011}, \emph{111}, 5179\relax
\mciteBstWouldAddEndPuncttrue
\mciteSetBstMidEndSepPunct{\mcitedefaultmidpunct}
{\mcitedefaultendpunct}{\mcitedefaultseppunct}\relax
\EndOfBibitem
\bibitem[Agranovich \latin{et~al.}(2012)Agranovich, Basko, and Rossa]{ref37}
Agranovich,~V.~M.; Basko,~D.~M.; Rossa,~G. C.~L. Efficient optical pumping of
  organic-inorganic heterostructures for nonlinear optics. \emph{Phys. Rev. B}
  \textbf{2012}, \emph{86}, 165204\relax
\mciteBstWouldAddEndPuncttrue
\mciteSetBstMidEndSepPunct{\mcitedefaultmidpunct}
{\mcitedefaultendpunct}{\mcitedefaultseppunct}\relax
\EndOfBibitem
\bibitem[King \latin{et~al.}(2012)King, Barbiellini, Moser, and
  Renugopalakrishnan]{ref38}
King,~C.; Barbiellini,~B.; Moser,~D.; Renugopalakrishnan,~V. \emph{Phys. Rev.
  B} \textbf{2012}, \emph{85}, 125106\relax
\mciteBstWouldAddEndPuncttrue
\mciteSetBstMidEndSepPunct{\mcitedefaultmidpunct}
{\mcitedefaultendpunct}{\mcitedefaultseppunct}\relax
\EndOfBibitem
\bibitem[Golovinskii(2014)]{ref39}
Golovinskii,~P.~A. \emph{Semiconductors} \textbf{2014}, \emph{48}, 760\relax
\mciteBstWouldAddEndPuncttrue
\mciteSetBstMidEndSepPunct{\mcitedefaultmidpunct}
{\mcitedefaultendpunct}{\mcitedefaultseppunct}\relax
\EndOfBibitem
\bibitem[Poddubny and Rodina(2016)Poddubny, and Rodina]{ref40}
Poddubny,~A.; Rodina,~A.~V. Nonradiative and radiative Forster energy transfer
  between quantum dots. \emph{ZhETF} \textbf{2016}, \emph{149}, 614\relax
\mciteBstWouldAddEndPuncttrue
\mciteSetBstMidEndSepPunct{\mcitedefaultmidpunct}
{\mcitedefaultendpunct}{\mcitedefaultseppunct}\relax
\EndOfBibitem
\bibitem[Hernandez-Martinez \latin{et~al.}(2014)Hernandez-Martinez, Govorov,
  and Demir]{add02}
Hernandez-Martinez,~P.~L.; Govorov,~A.~O.; Demir,~H.~V. \emph{J. of Phys. Chem.
  C} \textbf{2014}, \emph{118}, 4951--4958\relax
\mciteBstWouldAddEndPuncttrue
\mciteSetBstMidEndSepPunct{\mcitedefaultmidpunct}
{\mcitedefaultendpunct}{\mcitedefaultseppunct}\relax
\EndOfBibitem
\bibitem[Reich and Shklovskii(2016)Reich, and Shklovskii]{add03}
Reich,~K.~V.; Shklovskii,~B.~I. Exciton Transfer in Array of Epitaxially
  Connected Nanocrystals. \emph{ACS Nano} \textbf{2016}, \emph{10}, 10267\relax
\mciteBstWouldAddEndPuncttrue
\mciteSetBstMidEndSepPunct{\mcitedefaultmidpunct}
{\mcitedefaultendpunct}{\mcitedefaultseppunct}\relax
\EndOfBibitem
\bibitem[Kane(1957)]{ref41}
Kane,~E.~O. \emph{J. Phys. Chem. Solids} \textbf{1957}, \emph{1}, 249\relax
\mciteBstWouldAddEndPuncttrue
\mciteSetBstMidEndSepPunct{\mcitedefaultmidpunct}
{\mcitedefaultendpunct}{\mcitedefaultseppunct}\relax
\EndOfBibitem
\bibitem[Polkovnikov and Zegrya(1998)Polkovnikov, and Zegrya]{ref42}
Polkovnikov,~A.~S.; Zegrya,~G.~G. \emph{Phys. Rev. B} \textbf{1998}, \emph{58},
  4039\relax
\mciteBstWouldAddEndPuncttrue
\mciteSetBstMidEndSepPunct{\mcitedefaultmidpunct}
{\mcitedefaultendpunct}{\mcitedefaultseppunct}\relax
\EndOfBibitem
\bibitem[Zegrya and Samosvat(2007)Zegrya, and Samosvat]{ref43}
Zegrya,~G.~G.; Samosvat,~D.~M. \emph{J. Exp. Theor Phys.} \textbf{2007},
  \emph{104}, 951\relax
\mciteBstWouldAddEndPuncttrue
\mciteSetBstMidEndSepPunct{\mcitedefaultmidpunct}
{\mcitedefaultendpunct}{\mcitedefaultseppunct}\relax
\EndOfBibitem
\bibitem[Bir and Pikus(1974)Bir, and Pikus]{ref44}
Bir,~G.; Pikus,~G. \emph{Symmetry and Strain-induced Effects in
  Semiconductors}; A Halsted Press book; Wiley, 1974\relax
\mciteBstWouldAddEndPuncttrue
\mciteSetBstMidEndSepPunct{\mcitedefaultmidpunct}
{\mcitedefaultendpunct}{\mcitedefaultseppunct}\relax
\EndOfBibitem
\bibitem[Samosvat \latin{et~al.}(2015)Samosvat, Chikalova-Luzina, and
  Zegrya]{ref45}
Samosvat,~D.~M.; Chikalova-Luzina,~O.~P.; Zegrya,~G.~G. Nonradiative resonance
  energy transfer between semiconductor quantum dots. \emph{J. Exp. Theor
  Phys.} \textbf{2015}, \emph{121}, 76\relax
\mciteBstWouldAddEndPuncttrue
\mciteSetBstMidEndSepPunct{\mcitedefaultmidpunct}
{\mcitedefaultendpunct}{\mcitedefaultseppunct}\relax
\EndOfBibitem
\bibitem[Varshalovich \latin{et~al.}(1988)Varshalovich, Moskalev, and
  Khersonskii]{ref57}
Varshalovich,~D.~A.; Moskalev,~A.~N.; Khersonskii,~V.~K. \emph{Quantum Theory
  of Angular Momentum}; World Scientific: Singapore, 1988\relax
\mciteBstWouldAddEndPuncttrue
\mciteSetBstMidEndSepPunct{\mcitedefaultmidpunct}
{\mcitedefaultendpunct}{\mcitedefaultseppunct}\relax
\EndOfBibitem
\bibitem[Bateman and Erdelyi(1953)Bateman, and Erdelyi]{ref46}
Bateman,~H.; Erdelyi,~A. \emph{Higher Transcendental Functions}; McGraw-Hill:
  New York, 1953\relax
\mciteBstWouldAddEndPuncttrue
\mciteSetBstMidEndSepPunct{\mcitedefaultmidpunct}
{\mcitedefaultendpunct}{\mcitedefaultseppunct}\relax
\EndOfBibitem
\bibitem[Prudnikov \latin{et~al.}(2003)Prudnikov, Brychkov, and
  Marichev]{ref47}
Prudnikov,~A.; Brychkov,~Y.~A.; Marichev,~O.~I. \emph{Integrals and Series,
  Vol. 3: Special Functions: Additional Chapters}; Nauka: Moscow, 2003\relax
\mciteBstWouldAddEndPuncttrue
\mciteSetBstMidEndSepPunct{\mcitedefaultmidpunct}
{\mcitedefaultendpunct}{\mcitedefaultseppunct}\relax
\EndOfBibitem
\bibitem[Landau and Lifshitz(2005)Landau, and Lifshitz]{ref48}
Landau,~L.~D.; Lifshitz,~E.~M. \emph{Course of Theoretical Physics, Vol. 3:
  Quantum mechanics: Non-Relativistic Theory}; Butterworth-Heinemann: Oxford,
  2005\relax
\mciteBstWouldAddEndPuncttrue
\mciteSetBstMidEndSepPunct{\mcitedefaultmidpunct}
{\mcitedefaultendpunct}{\mcitedefaultseppunct}\relax
\EndOfBibitem
\bibitem[Pantell and Puthoff(1969)Pantell, and Puthoff]{ref49}
Pantell,~R.; Puthoff,~G. \emph{Fundamentals of Quantum Electronics}; Plenum:
  New York, 1969\relax
\mciteBstWouldAddEndPuncttrue
\mciteSetBstMidEndSepPunct{\mcitedefaultmidpunct}
{\mcitedefaultendpunct}{\mcitedefaultseppunct}\relax
\EndOfBibitem
\bibitem[Konyshev and Burstein(1968)Konyshev, and Burstein]{ref50}
Konyshev,~V.~P.; Burstein,~A.~I. \emph{Teor. Eksp. Khim.} \textbf{1968},
  \emph{4}, 192\relax
\mciteBstWouldAddEndPuncttrue
\mciteSetBstMidEndSepPunct{\mcitedefaultmidpunct}
{\mcitedefaultendpunct}{\mcitedefaultseppunct}\relax
\EndOfBibitem
\bibitem[Samosvat \latin{et~al.}(2013)Samosvat, Evtikhiev, Shkol'nik, and
  Zegrya]{ref51}
Samosvat,~D.~M.; Evtikhiev,~V.~P.; Shkol'nik,~A.~S.; Zegrya,~G.~G. On the
  lifetime of charge carriers in quantum dots at low temperatures.
  \emph{Semiconductors} \textbf{2013}, \emph{47}, 22\relax
\mciteBstWouldAddEndPuncttrue
\mciteSetBstMidEndSepPunct{\mcitedefaultmidpunct}
{\mcitedefaultendpunct}{\mcitedefaultseppunct}\relax
\EndOfBibitem
\bibitem[Inoshita and Sakaki(1992)Inoshita, and Sakaki]{ref52}
Inoshita,~T.; Sakaki,~H. \emph{Phys. Rev. B} \textbf{1992}, \emph{46},
  7260\relax
\mciteBstWouldAddEndPuncttrue
\mciteSetBstMidEndSepPunct{\mcitedefaultmidpunct}
{\mcitedefaultendpunct}{\mcitedefaultseppunct}\relax
\EndOfBibitem
\bibitem[Wang \latin{et~al.}(1994)Wang, Fafard, Leonard, Bowers, Merz, and
  Petroff]{ref53}
Wang,~G.; Fafard,~S.; Leonard,~D.; Bowers,~J.~E.; Merz,~J.~L.; Petroff,~P.~M.
  \emph{Appl. Phys. Lett} \textbf{1994}, \emph{64}, 2815\relax
\mciteBstWouldAddEndPuncttrue
\mciteSetBstMidEndSepPunct{\mcitedefaultmidpunct}
{\mcitedefaultendpunct}{\mcitedefaultseppunct}\relax
\EndOfBibitem
\bibitem[Asryan and Suris(2004)Asryan, and Suris]{ref54}
Asryan,~L.~V.; Suris,~R.~A. Theory of threshold characteristics of
  semiconductor quantum dot lasers. \emph{Semiconductors} \textbf{2004},
  \emph{38}, 1\relax
\mciteBstWouldAddEndPuncttrue
\mciteSetBstMidEndSepPunct{\mcitedefaultmidpunct}
{\mcitedefaultendpunct}{\mcitedefaultseppunct}\relax
\EndOfBibitem
\bibitem[Heitz \latin{et~al.}(1997)Heitz, Veit, Ledentzov, Hoffman, Bimberg,
  Ustinov, Kop'ev, and Alferov]{ref55}
Heitz,~R.; Veit,~M.; Ledentzov,~N.~N.; Hoffman,~A.; Bimberg,~D.;
  Ustinov,~V.~M.; Kop'ev,~P.~S.; Alferov,~Z. \emph{Phys. Rev. B} \textbf{1997},
  \emph{56}, 10435\relax
\mciteBstWouldAddEndPuncttrue
\mciteSetBstMidEndSepPunct{\mcitedefaultmidpunct}
{\mcitedefaultendpunct}{\mcitedefaultseppunct}\relax
\EndOfBibitem
\bibitem[Baranov \latin{et~al.}(2000)Baranov, Davydov, Ren, Sugoou, and
  Masumoto]{ref56}
Baranov,~A.~V.; Davydov,~V.; Ren,~H.~W.; Sugoou,~S.; Masumoto,~Y. \emph{Journal
  of Luminescence} \textbf{2000}, \emph{87}, 503\relax
\mciteBstWouldAddEndPuncttrue
\mciteSetBstMidEndSepPunct{\mcitedefaultmidpunct}
{\mcitedefaultendpunct}{\mcitedefaultseppunct}\relax
\EndOfBibitem
\bibitem[Monguzzi \latin{et~al.}(2008)Monguzzi, Tubino, and Meinardi]{ref62}
Monguzzi,~A.; Tubino,~R.; Meinardi,~F. Unconversipon-indused delayed
  fluorescence in multicomponent organic systems: Role of {Dexter} energy
  transfer. \emph{Phys. Rev. B} \textbf{2008}, \emph{77}, 155122\relax
\mciteBstWouldAddEndPuncttrue
\mciteSetBstMidEndSepPunct{\mcitedefaultmidpunct}
{\mcitedefaultendpunct}{\mcitedefaultseppunct}\relax
\EndOfBibitem
\bibitem[Hoffman \latin{et~al.}(2014)Hoffman, Choi, and Kamat]{add04}
Hoffman,~J.~B.; Choi,~H.; Kamat,~P.~V. \emph{J. Phys. Chem. C} \textbf{2014},
  \emph{118}, 18453–18461\relax
\mciteBstWouldAddEndPuncttrue
\mciteSetBstMidEndSepPunct{\mcitedefaultmidpunct}
{\mcitedefaultendpunct}{\mcitedefaultseppunct}\relax
\EndOfBibitem
\bibitem[Mork \latin{et~al.}(2014)Mork, Weidman, Prins, and Tisdale]{add05}
Mork,~A.~J.; Weidman,~M.~C.; Prins,~F.; Tisdale,~W.~A. \emph{J Phys. Chem. C}
  \textbf{2014}, \emph{118}, 13920-- 13928\relax
\mciteBstWouldAddEndPuncttrue
\mciteSetBstMidEndSepPunct{\mcitedefaultmidpunct}
{\mcitedefaultendpunct}{\mcitedefaultseppunct}\relax
\EndOfBibitem
\bibitem[Li \latin{et~al.}(2006)Li, Ma, Wang, and Su]{ref58}
Li,~Y.; Ma,~Q.; Wang,~X.; Su,~X. Fluorescence resonance energy transfer between
  two quantum dots with immunocomplexes of antigen and antibody as a bridge.
  \emph{Luminescence} \textbf{2006}, \emph{22}, 60\relax
\mciteBstWouldAddEndPuncttrue
\mciteSetBstMidEndSepPunct{\mcitedefaultmidpunct}
{\mcitedefaultendpunct}{\mcitedefaultseppunct}\relax
\EndOfBibitem
\bibitem[Xin-YanWeng \latin{et~al.}(2005)Xin-YanWeng, Ma, Li, Li, Su, and
  Qin-HanJin]{ref59}
Xin-YanWeng,; Ma,~Q.; Li,~Y.-B.; Li,~B.; Su,~X.-G.; Qin-HanJin, \emph{Canadian
  Journal of Analytical Sciences and Spectroscopy} \textbf{2005}, \emph{50},
  141\relax
\mciteBstWouldAddEndPuncttrue
\mciteSetBstMidEndSepPunct{\mcitedefaultmidpunct}
{\mcitedefaultendpunct}{\mcitedefaultseppunct}\relax
\EndOfBibitem
\bibitem[Schobel \latin{et~al.}(1999)Schobel, Egelhaaf, Brecht, Oelkrug, and
  Gauglitz]{ref60}
Schobel,~U.; Egelhaaf,~H.~J.; Brecht,~A.; Oelkrug,~D.; Gauglitz,~G.
  \emph{BioconjugateChem.} \textbf{1999}, \emph{10}, 1107\relax
\mciteBstWouldAddEndPuncttrue
\mciteSetBstMidEndSepPunct{\mcitedefaultmidpunct}
{\mcitedefaultendpunct}{\mcitedefaultseppunct}\relax
\EndOfBibitem
\bibitem[Liu \latin{et~al.}(2008)Liu, Zhang, and Jian-HaoWang]{ref61}
Liu,~T.-C.; Zhang,~H.-L.; Jian-HaoWang, \emph{Anal. Bioanal. Chem.}
  \textbf{2008}, \emph{391}, 2819\relax
\mciteBstWouldAddEndPuncttrue
\mciteSetBstMidEndSepPunct{\mcitedefaultmidpunct}
{\mcitedefaultendpunct}{\mcitedefaultseppunct}\relax
\EndOfBibitem
\end{mcitethebibliography}
\providecommand{\latin}[1]{#1}
\providecommand*\mcitethebibliography{\thebibliography}
\csname @ifundefined\endcsname{endmcitethebibliography}
  {\let\endmcitethebibliography\endthebibliography}{}

\end{document}